\documentclass[iop]{emulateapj}
\usepackage{amsmath}
\usepackage{amsbsy}
\usepackage{amssymb}
\usepackage{graphicx}
\usepackage{calrsfs}
\usepackage{natbib}

\usepackage{hyperref}
\DeclareMathAlphabet\mathbfcal{OMS}{cmsy}{b}{n}

\begin{document}

\def\b{\boldsymbol}
\def\p{\partial}
\def\vt{\vartheta}
\def\e{\epsilon}
\def\ts{\textstyle}
\def\k{\varkappa}
\def\d{\delta}
\def\be{\begin{equation}}
\def\ee{\end{equation}}

\title{Synchro-curvature radiation of charged particles \\ in the strong curved magnetic fields}

\author{S.R.~Kelner}
\affiliation
{Max-Planck-Institut f\"ur Kernphysik,
Saupfercheckweg 1, D-69117 Heidelberg, Germany}
\affiliation{Research Nuclear University (MEPHI), Kashirskoe shosse 31,
115409 Moscow, Russia}
\email{Stanislav.Kelner@mpi-hd.mpg.de}

\author{A.Yu.~Prosekin}
\affiliation{Max-Planck-Institut f\"ur Kernphysik,
Saupfercheckweg 1, D-69117 Heidelberg, Germany}
\email{Anton.Prosekin@mpi-hd.mpg.de}

\author{F.A.~Aharonian}
\affiliation{Dublin Institute for Advanced Studies, 31 Fitzwilliam Place,
Dublin 2, Ireland}
\affiliation
{Max-Planck-Institut f\"ur Kernphysik,
Saupfercheckweg 1, D-69117 Heidelberg, Germany}
\email{Felix.Aharonian@mpi-hd.mpg.de}

\date{\today}

\begin{abstract}
It is generally believed that the radiation of relativistic particles in a curved magnetic field proceeds in either the
synchrotron or the curvature radiation modes. In this paper we show that in strong curved magnetic fields a
significant fraction of the energy of relativistic electrons can be radiated away in the intermediate, the so-called
synchro-curvature regime. Because of the persistent change of the trajectory curvature, the radiation varies with the
frequency of particle gyration. While this effect can be ignored in the synchrotron and curvature regimes, the
variability plays a key role in the formation of the synchro-curvature radiation. Using the Hamiltonian formalism,
we find that the particle trajectory has the form of a helix wound around the drift trajectory. This allows us to
calculate analytically the intensity and energy distribution of prompt radiation in the general case of magnetic
bremsstrahlung in the curved magnetic field. We show that the transition to the limit of the synchrotron and
curvature radiation regimes is determined by the relation between the drift velocity and the component of the
particle velocity perpendicular to the drift trajectory. The detailed numerical calculations, which take into account
the energy losses of particles, confirm the principal conclusions based on the simplified analytical treatment of the
problem, and allow us to analyze quantitatively the transition between different radiation regimes for a broad range
of initial pitch angles. These calculations demonstrate that even very small pitch angles may lead to significant
deviations from the spectrum of the standard curvature radiation when it is formally assumed that a charged particle
moves strictly along the magnetic line. We argue that in the case of realization of specific configurations of the
electric and magnetic fields, the gamma-ray emission of the pulsar magnetospheres can be dominated by the
component radiated in the synchro-curvature regime.
\end{abstract}

 
\maketitle

\section{Introduction}

The energy distribution and intensity of radiation of a
charged particle moving in the magnetic field with a relativistic
speed is determined by the curvature of its trajectory. The
trajectory in the curved magnetic field has a structure of a helix
wound along the so-called \textit{drift trajectory}, which describes the
particle motion averaged over the fast oscillations. Thus, the
curvature of the trajectory is formed by the curvature of the
helix itself and the curvature of the drift trajectory; they are
determined by the strength and the curvature of the magnetic
field, respectively. If the main contributor is the curvature of
the helix (i.e., the bending of the drift trajectory can be
neglected), the radiation is defined by the strength of the
magnetic field. This is the case of the synchrotron radiation. On
the other hand, if the helix is stretched (i.e., the particle moves
along the drift trajectory with a little wobbling around it), the
main contribution to the radiation is due to the curvature of the
drift trajectory. This is the case of the curvature radiation,
which is determined by curvature of the drift trajectory
\footnote{Note that usually in the literature the curvature radiation is considered as
radiation when particles move along the magnetic line. Strictly speaking, this is
a formal but, in fact, not a correct assumption because a particle can move only
along a drift trajectory. However, because the drift trajectory is close to the
magnetic line in the strong magnetic field, this assumption does not lead to a
significant inaccuracy.}.
Obviously, the synchrotron and curvature radiation regimes are
approximations that work with a high precision as long as the
curvatures of the helix and the drift trajectory differ
significantly. However, when they are comparable their joint
action may result in features that are significantly different from
those described by the standard synchrotron and curvature
radiation formalism. These two contributions to the curvature
can reinforce or extinguish each other, and therefore cause a
significant variability of the curvature of the particle trajectory.
It results, in turn, in a radiation that varies with the frequency of
the particle gyration. The observer detects the time-averaged
energy distribution of this promptly variable radiation, which
appears rather different from the spectra of emission produced
in the synchrotron or curvature radiation regimes.

A few attempts have been made in the past to study this
intermediate regime of radiation, which we, following \cite{Cheng1996}, 
call synchro-curvature radiation. However the
approaches and assumptions used in the previous treatments of
the problem \citep{Cheng1996,Harko2002} are
generally not self-consistent. For example, in the paper of
\cite{Cheng1996}, where the particle motion is
introduced a priori (“by hand”), the square of the velocity is
not conserved. The more recent results of \cite{Harko2002} also seem not correct. 
In particular, the acceleration in
this work is not variable as it follows from the solution of
equations that describe the particle trajectories (compare the
formula given by Equation~(5) in their paper and the expression
given by Equation~(\ref{cur32}) in the present work). On the other
hand, we cannot fail to mention the apparently underrated work
of \cite{Sobolev2006}, where the author has thoroughly treated the
problem of radiation of particles in the magnetic field. Some
principal results and conclusions of this paper are in a good
agreement with the results of Section~\ref{sec:traj} of the present work,
although they have been obtained using a rather different
computational approach to the problem. However, these results
are of rather theoretical interest because the energy losses of
particles are ignored.

In this work, using the Hamiltonian formalism, we solve the
equations of the motion of a charged particle in the azimuthally
symmetric magnetic field. This allows us to self-consistently
describe the local properties of the particle trajectory with a
time-dependent curvature, and calculate the time-averaged
spectrum of radiation in an arbitrary magnetic field. The
analytical solutions derived in an approximation that ignores
energy losses of electrons provide a useful tool for understanding
of distinct feature of the synchro-curvature radiation,
and analyzing the conditions of its transition to the limits of
synchrotron and curvature regimes.

On the other hand, for the quantitative description of
propagation and radiation of particles, one has to treat the
energy losses of electrons properly. For this reason, we have
performed numerical integration of the equations of motion in
the dipole magnetic field, taking into account the radiation
reaction force. The dipole magnetic field can be used as a good
approximation for the strong magnetic field in the vicinity of
compact objects like neutron stars. The results of numerical
calculations allow us to study the evolution of radiation for
different initial pitch-angles. These results demonstrate that
even a small initial deflection of the motion of electrons from
the drift trajectory could lead to dramatic deviations from the
radiation features expected in the conventional curvature
radiation regime. The question of characteristic values of
pitch-angles in the pulsar magnetospheres remains an open
issue. The answer can be obtained if one brings the electron
acceleration into consideration. However this requires a
knowledge of the relative arrangement of the electric and
magnetic fields. We will discuss this issue in the context of the
acceleration of electrons in the simple model of the
electric field.

This paper has the following structure. In Section~\ref{sec:traj} we
analytically solve the equations of the motion of a particle in
the magnetic field of a constant curvature. Using the curvature
of the trajectory, in Section~\ref{sec:inten} we derive analytical expressions
for the radiation spectrum of synchro-curvature radiation.
Subsequent sections deal with a numerical treatment of the
particle radiation in some specific configurations that are of
astrophysical interest. Section~\ref{sec:num}, as well as Appendices~\ref{sec:accel} and 
\ref{sec:appB}, describe relevant formalism for the particle motion and its
radiation. In Section~\ref{sec:implic} we discuss the results of numerical
calculations, namely the radiation spectra and regimes of
radiation for different initial conditions. In Section~\ref{sec:elec} and
Appendix~\ref{sec:maxen} we investigate the impact of the combined action
of electric and magnetic fields on the particle acceleration and
radiation. Finally, in Section~\ref{sec:dis} we summarize the main results
and conclusions.

\section{Particle trajectories at small pitch angles}\label{sec:traj}

The motion of a relativistic charged particle in a strong
magnetic field is accompanied by a radiation that in general
terms can be called magnetic bremsstrahlung. While moving in
a homogeneous field, a charged relativistic particle gradually
radiates away its energy. However, the component of its
velocity, which is parallel to the magnetic field, $v_{\parallel}^{}$, remains
unchanged. This statement is obvious for $v_{\parallel}^{}=0$, therefore in
an arbitrary coordinate system, which moves along (or
opposite) the magnetic field, we have $v_{\parallel}^{}={\rm const}$. Thus,
\be\label{in1}
\frac{c^2}{v_{\parallel}^2}-1=\frac{p_\perp^2+m^2c^2}{p_{\parallel}^2} = {\rm
const}\,,
\ee
where $p_{\parallel}^{}$ and $p_{\perp}^{}$ are the parallel and perpendicular components
of the momentum, respectively. This equation allows us to find
the energy of the particle after the radiative damping of the
perpendicular component of motion. We denote the initial and
final values of the energy, and the parallel and perpendicular
components of the momentum as$E$, $p_{\parallel}^{} $,
$p_{\perp}^{}$, and $E'$, $p'_{\parallel}$,
$p'_{\perp}=0$, respectively. From Equation (\ref{in1}) it follows that
\be\label{in2}
p'_{\parallel}=p_{\parallel}^{}\,\frac{mc}{\sqrt{p_{\perp}^2+m^2c^2}}\,,
\qquad E'=E\,\frac{mc}{\sqrt{p_{\perp}^2+m^2c^2}}\,.
\ee

If the initial perpendicular momentum is nonrelativistic,
$p_{\perp}^{}\ll mc$, then $E'\approx E$ (i.e., the damping of the perpendicular
motion does not have an effect on the particle energy).
However, for $mc\ll p_{\perp}^{}\ll p_{\parallel}^{}$, we have $E'\ll E$ (i.e., the
particle loses almost all its energy during the damping of the
perpendicular motion). For instance, for a particle with the
initial Lorentz factor $\gamma$ and the initial pitch angle
$\alpha=100/\gamma\ll 1$, the final Lorentz factor is reduced by two
orders of magnitude, $\gamma'=\gamma/100$. At $p_{\perp}^{}\gg mc$, the momenta
$p_{\parallel}^{}$ and $p_{\perp}^{}$ vary similarly with time because of the constant ratio
$p_{\perp}^{}/p_{\parallel}^{}={\rm const}$ (see Equation (\ref{in1})). The constancy of the ratio
$p_{\perp}^{}/p_{\parallel}^{}$ has a simple explanation. The ultrarelativistic particle
emits photons in the direction of its momentum $\b p$ within the
angle $\lesssim1/\gamma$, therefore in the case of $1/\gamma\ll p_{\perp}^{}/p_{\parallel}^{}$, the recoil
momentum is directed opposite to the vector $\b p$. No matter how
small the pitch angle is, as long as it is much larger than $1/\gamma$,
the particle will lose a considerable amount of its energy.

In the homogeneous field, the particle, after losing the
perpendicular component of its momentum, continues its
motion along the magnetic field line. The situation is quite
different when the field is curved. If the velocity vector is
parallel to the field line initially, the acceleration $\b a\big|_{t=0}=0$.
Correspondingly the curvature radius\footnote{Note that at the motion with a constant absolute speed, the curvature radius by definition is equal to $R_c=v^2/|\b a|$. }
$R_c\big|_{t=0}=\infty$, thus no
radiation is emitted at $t=0$. In the curved magnetic field, a
rectilinear uniform motion is not possible, so the acceleration
and radiation of the particle are unavoidable. For all trajectories
located close to the drift trajectory, the acceleration and
curvature radius appear to be time-dependent. This implies that
the radiation also should vary with time, thus for calculation of
the measurable characteristics of radiation one should average
the relevant distributions over time.

We start with a simple model of particle motion that allows
exact solutions. Let us consider the case when the magnetic
field has the same symmetry as the field of an infinitely long
straight wire. Then, in the cylindrical system of coordinates,
$(r,\,\phi,\,z)$, the field only has an azimuthal component; it only
depends on the distance to the axis of symmetry, $r$, but not on $z$
and $\phi$. We denote by $(\b e_r,\b e_\phi,\,\b e_z)$ a moving system of unit
vectors linked to the point $(r,\,\phi,\,z)$. The magnetic field can be
expressed as $\b B=B(r)\,\b e_{\phi}^{}$. The field lines are circles of a radius
$r$, which means that at a particular distance from the axis of
symmetry the magnetic field has a constant curvature. Particles
in such a field can move along trajectories close to the circular
field lines.

The equation that describes the particle motion is
\be\label{cur1}
m\gamma\dot{\b v}=\frac ec\,(\b v\times\b B)\,.
\ee
Let us find a solution for which $r=r_0={\rm const}$, and the
velocity
\be\label{cur2}
\b v=v_\phi\b e_\phi+v_z \b e_z\, .
\ee
Note that the components $v_\phi$ and $v_z$ are constants. The vector $\b e_\phi$
is uniformly rotating with an angular frequency $\omega=v_\phi/r_0$,
therefore the derivative $d\b e_\phi/dt= -v_\phi\b e_r/r$ and the vector
$\b e_z ={\rm const}$. Substituting Equation (\ref{cur2}) into (\ref{cur1}), we find
\be\label{cur3}
m\gamma \frac{v_\phi^2}{r_0}\,\b e_r=\frac ec\,v_z B\,\b e_r\,.
\ee
From here it follows that
\be\label{cur4}
\frac{v_z}{v_\phi}=\frac{mcv_\phi\gamma}{eB\,r_0}\,.
\ee
Writing the velocity components in the form $v_z=v\sin\alpha$,
$v_\phi=v\cos\alpha$ (the angle is counted from the $(x,y)$ plane, i.e.,
$-\pi/2\le \alpha \le \pi/2$), and expressing the velocity $v$ via the
particle Lorentz factor $\gamma$, Equation (\ref{cur4}) can be presented in
the following form

\be\label{cur5}
\frac{\sin\alpha}{\cos^2\alpha}=\frac{mc^2\gamma}{eB\,r_0}\,
\sqrt{1-1/\gamma^2
} \, .
\ee
Apparently, for any given $\gamma$ this equation defines the angle $\alpha$.
Equation (\ref{cur1}) indeed has solutions in the form of Equation (\ref{cur2}),
which describes the helicoidal motion depending on the
parameters $\gamma$ and $r_0$. Note that a motion with strictly circular
trajectories is not possible. Thus the particle trajectory can be
interpreted as a motion along the field line with a simultaneous
drift in the perpendicular direction. The positive and negative
charges drift in opposite directions.

The condition for a small step in the helix, $|\alpha| \ll 1$, requires
\be\label{cur6}
1\ll \gamma \ll \frac{|e|B\,r_0}{mc^2}=6\times 10^{13} \Big(\frac{B}{10^{11}
\;{\rm G}}\Big) \Big( \frac{r_0}{10\;{\rm km}}\Big)\,.
\ee
Note that for typical pulsar parameters this condition is readily
satisfied, and thus
\be\label{cur7}
\alpha=\frac{mc^2\gamma}{eB\,r_0}\,; \qquad |\alpha|\ll 1 \,.
\ee

Under the condition of Equation (\ref{cur6}), the curvature radius $R_c$
of the trajectory of the particle located at the distance $r_0$ does
not depend on the particle energy; it practically coincides with
the curvature radius of the field line $r_0$. Note that this
conclusion is correct for any configuration of the magnetic
field. Indeed, the short segments of the field line can be treated
as a circle, thus when satisfying the condition of Equation (\ref{cur6})
the particle can move at this segment strictly along the drift
trajectory.

For the calculation of the prompt emission of the particle
moving along drift trajectory, the curvature of the field line can
be used instead of the curvature of the trajectory because
$R_c\approx r_0$. However, one cannot use the magnetic field line
instead of the drift trajectory as a trajectory for the particle. As
shown below, the characters of the motion of the particle
initially moving along the magnetic field line and the drift
trajectory are considerably different. Therefore, the interpretation
of the curvature radiation as the radiation of the particle
moving along the magnetic line is not correct. The strict
formulation of the curvature radiation corresponds to the
radiation of the particle moving along the drift trajectory. This
is the definition that we will use for the curvature radiation
throughout this paper. 

The intensity of the curvature radiation is defined as
\be\label{cur8}
I_{\rm curv}=\frac{2e^2c}{3}\,\frac{\gamma^4}{R_c^2}
\approx \frac{2e^2c}{3}\,\frac{\gamma^4}{r_0^2}\,.
\ee
When the particle moves close to, but not exactly along the
drift trajectory, the intensity of radiation can deviate from
Equation (\ref{cur8}). Therefore one should carefully investigate the
trajectories in the vicinity of the drift trajectories. For this
purpose, we invoke the Hamiltonian formalism. The vector
potential can be taken in the form $A_x=0$, $A_y=0$, $A_z=A(r)$,
therefore the azimuthal component of the field
$B(r)=-dA/dr$. In the cylindrical coordinates, the Hamilton
function can be written as
\be\label{cur9}
H=c\,\sqrt{P_r^2+\frac{P_\phi^2}{r^2}+(P_z-\frac ec\,A(r))^2+m^2c^2}\,,
\ee
where $P_r$, $P_\phi$, and $P_z$ are the generalized momenta corresponding
to the coordinates $r$, $\phi$ and $z$, respectively. Because of the
azimuthal symmetry of the magnetic field and its homogeneity
along the $z$ axis, $\phi$ and $z$ are cyclic coordinates. Therefore $P_\phi$
and $P_z$, as well as the energy, are integrals of motion. The
presence of three integrals of motion allows us to reduce the
problem to the one-dimensional case. Below we will limit the
treatment to the case of ultrarelativistic particles and small
pitch-angles.

For given $P_\phi$ and $P_z$, the radial motion can be considered, as
it follows from Equation (\ref{cur9}), as a motion in the field with the
effective potential:

\be\label{cur10}
U_{\rm eff}(r)=\frac{P_\phi^2}{r^2}+(P_z-\frac ec\,A(r))^2\,.
\ee
The first and second derivatives of this potential are
\be\label{cur11}
U'_{\rm eff}(r)=-\frac{2P_\phi^2}{r^3}+\frac{2e}{c}B(r)
(P_z- \frac ec\,A(r))\,,
\ee
and
\be\label{cur12}
U''_{\rm eff}(r)=\frac{6P_\phi^2}{r^4}+\frac{2e}{c}B'(r)
(P_z- \frac ec\,A(r))+\frac{2e^2}{c^2}\,B^2(r)\,.
\ee
Assuming that at $r=r_0$, the derivative $U'_{\rm eff}(r_0)=0$, we can
write 
\be\label{cur13}
P_z-\frac{e}{c}A(r_0)=\frac{cP_\phi^2}{r_0^3 e B(r_0)}\,,
\ee
\be\label{cur14}
U''_{\rm eff}(r_0)=\frac{2P_\phi^2}{r_0^4}\left(3+\frac{r_0B'(r_0)}{B(r_0)}
\right) +\frac{2e^2}{c^2}\,B^2(r_0)\,.
\ee

If $B$ decreases with $r$ as a power-law, $B\sim 1/r^\d$, then for
$\d <3$ both terms in Equation~(\ref{cur14}) are positive, and therefore
$U_{\rm eff}$ has a minimum at $r_0$. Furthermore, by assuming that the
condition of Equation~(\ref{cur6}) is fulfilled, we can keep only the last
term in Equation~(\ref{cur14}).

For integrating the equations of motion, we apply the
standard method that is used in the theory of small oscillations.
In the vicinity of the minimum point, the function $U_{\rm eff}(r)$ can
be approximated by a parabola:
\be\label{cur15}
U_{\rm eff}(r)=P_0^2+m^2\omega_c^2(r-r_0)^2\,,
\ee
where
\be\label{cur16}
\omega_c^2=\frac{1}{2m^2}U''_{\rm eff}(r_0)\approx
\left(\frac{eB(r_0)}{mc} \right)^2\,,
\ee
\be\label{cur17}
P_0^2=U_{\rm eff}(r_0)=
\frac{P_\phi^2}{r_0^2}\left(1+\frac{P_\phi^2}{r_0^4m^2\omega_c^2} \right).
\ee
Here $\omega_c$ is the cyclotron frequency.

In these denotations, the Hamilton function can be written in
the form
\be\label{cur18}
H=c\sqrt{P_r^2+m^2\omega_c^2(r-r_0)^2+P_0^2+m^2c^2}\, ,
\ee
and correspondingly the equations of motion are
\be\label{cur19}
\dot r=\frac{\p\tilde H}{\p P_r}=\frac{c^2P_r}{E}\,,\quad
\dot P_r=-\frac{\p\tilde H}{\p r}=-\frac{m^2c^2\omega_c^2}{E}\,(r-r_0)\,,
\ee
where $E$ is the particle energy. The solution of these Equations
gives
\be\label{cur20}
r=r_0-\rho\cos\frac{\omega_ct}{\gamma}\,,\quad
P_r=m\rho\omega_c\sin\frac{\omega_ct}{\gamma}\,,
\ee
where $\gamma=E/(mc^2)$ is the Lorentz factor of the particle, and $\rho$
is an arbitrary constant\footnote{Note that here the origin of the coordinate system is arbitrary, therefore instead of $t$ one can use $t-t_0$, where $t_0$ is the second arbitrary constant. Below
we omit the arbitrary constants that can be added to $\phi$ and $z$.} related to the particle energy
through the relation
\be\label{cur20a}
E=c\sqrt{P_0^2+m^2\rho^2\omega_c^2+m^2c^2}\,.
\ee
The condition of the applicability of Equation (\ref{cur20}) is the
smallness of $\rho$ compared to $r_0$, as well as to the characteristic
linear scale on which the magnetic field is changed
significantly.

Now we should find the time-dependencies of other
coordinates. We note that in this region
\be\label{cur21}
\dot\phi=\frac{\p H}{\p P_\phi}=\frac{P_\phi}{m\gamma}\,\frac{1}{r^2}
\approx \frac{P_\phi}{m\gamma}\,\frac{1}{r_0^2}
\left(1-\frac{2(r-r_0)}{r_0}\right) \,.
\ee
Using Equation (\ref{cur20}), we find
\begin{eqnarray}
\dot\phi&=&\frac{P_\phi}{m\gamma}\,\frac{1}{r_0^2}
\left(1+\frac{2\rho}{r_0}\cos\frac{\omega_ct}{\gamma}\right),\label{cur22}\\
\phi&=&\frac{P_\phi}{m\gamma}\,\frac{1}{r_0^2}
\left(t+\frac{2\rho\gamma}{\omega_c
r_0}\sin\frac{\omega_ct}{\gamma}\right).\label{cur23}
\end{eqnarray}
Substituting in
\be\label{cur24}
\dot z=\frac{\p H}{\p P_z}=\frac{c^2}{E}\,\left(P_z-\frac ec A(r)\right)
\ee
the following approximate expression  for $A(r)$,
\be\label{cur24a}
A(r)=A\big(r_0+(r-r_0)\big)\approx A(r_0)-B(r_0)\cdot(r-r_0) \, ,
 \ee
and using Equations~(\ref{cur13}) and (\ref{cur20}), we find 
\begin{eqnarray}
\dot z&=&\frac{P_\phi^2}{\gamma\omega_c m^2r_0^3}-\frac{\rho\,\omega_c}{\gamma}
\cos\frac{\omega_ct}{\gamma}\,,\label{cur25}\\
z&=&\frac{P_\phi^2}{\gamma\omega_c m^2r_0^3}\,t-\rho
\sin\frac{\omega_ct}{\gamma}\,.\label{cur26}
\end{eqnarray}

In order to make the analytical expressions more convenient
to work with it is useful to use, instead of $P_\phi$ and $\rho$, the
following parameters
\be\label{cur27}
\beta_\parallel^{}= \frac{P_\phi}{mc\gamma r_0}\,,\quad
\beta_\perp^{}=\frac{\rho\,\omega_c}{\gamma c} \, ,
\ee
and introduce one more parameter, 
\be\label{cur28}
\beta_D^{}=\frac{P_\phi^2}{c\gamma\omega_c m^2r_0^3}=\frac{c\gamma
\beta_\parallel^2}
{r_0\,\omega_c}\,.
\ee

Since $P_\phi$ is the projection of the particle momentum on the
axis $z$, $\beta_\parallel^{}$ is the parallel component of the velocity relative to
the magnetic field, while $\beta_\perp^{}$ is the component of the velocity
perpendicular to the drift trajectory, and $\beta_D^{}$ is the drift velocity
(all in units of $c$).

For the new variables we find
\be\label{cur29a}
 \dot r/c=\beta_\perp^{} \sin\tau\,,\quad \dot z/c=\beta_D^{}-\beta_\perp^{}
\cos\tau\,,
 \ee
 \be\label{cur29b}
\dot\phi= \frac{\beta_\parallel^{}c}{r_0}
\left(1+\frac{2\beta_\perp^{}\beta_D^{}}
{\beta_{\parallel}^2}\,\cos\tau\right),
\ee
where $\tau=\omega_c t/\gamma$, and the azimuthal component of the velocity
is
\be\label{cur29c}
v_\phi^{}=r\dot\phi=c\beta_{\parallel}^{}\left(1+\frac{\beta_\perp^{}\beta_D^{}}
{\beta_{\parallel}^2}\,\cos\tau\right).
\ee

For $\beta_\perp^{}=0$, this solution coincides with the previously
obtained exact solution given by Equation~(\ref{cur4}), which describes
the motion along a drift trajectory (in this case, along the
helicoidal line). Note that in this case the quantity $\beta_D$ plays the
role of $v_z$ and describes the relativistic curvature drift. The same
trajectory can be found if we average Equations (\ref{cur29a})-(\ref{cur29c}) over
time. Then this solution can be interpreted as the motion along
the helix around the drift trajectory. In the general case of an
arbitrary magnetic field, motion strictly along drift trajectory is
possible only locally, as long as the curvature of the field line is
approximately constant. At $\beta_\perp^{}\ne 0$, the solution is 
approximate; it is correct when the following condition is
fulfilled:

\be\label{cur30}
\beta_\perp^{}\ll \beta_\parallel^{}\,,\qquad
\beta_D^{}\ll \beta_\parallel^{}\,.
\ee
The relation between $\beta_\perp^{}$ and $\beta_D^{}$ can be arbitrary. Since 
$\beta_\parallel^{}\approx 1$, the second condition in Equation~(\ref{cur30}) is equivalent to Equation~(\ref{cur6}). The same requirement can be formulated in terms of the
smallness of the Larmor radius compared to the curvature
radius of the field line.

Note that Equation~(\ref{cur24a}) does not include the term
corresponding to the so-called gradient drift. Formally, the
gradient drift could be taken into account if we add to
Equation~(\ref{cur24a}) the next (quadratic) term of expansion. However
our study is essentially linear, so the inclusion of the gradient
drift could hardly be justified because it would exceed the
accuracy of the approach.

For the chosen new variables, the velocity and its square are
\be
\b v=\dot r\b e_r+r\dot\phi\b e_\phi+\dot z\b e_z\,,
\ee
and
\be
\b v^2=c^2\left(\beta_\parallel^2+\beta_\perp^2+\beta_D^2 \right)\,.
\ee
It is important to note that $\b v^2$ does not depend on $t$. In order to
find the acceleration, we use the following equation
\be\label{cur31}
\b a=\frac{e}{mc\gamma}\,(\b v\times\b B)=
\frac{\omega_c}{\gamma}\,(\dot r\b e_z-\dot z\b e_r)\,,
\ee
where it is taken into account that the field has only one
(azimuthal) component. The square of acceleration is
\be\label{cur32}
\b a^2=a_0^2\left(1-2\eta\cos\tau+\eta^2\right),
\ee
where
\be
a_0^{}=c^2\beta_\parallel^2/r_0\, \,  {\rm and}  \, \, \eta=\beta_\perp/\beta_D.
\ee
This equation can also be derived through the consideration of
kinematics. Indeed, taking into account in the equation
\be\label{cur33}
\b a=\ddot{\b r} =(\ddot r-r\dot\phi^2)\b e_r+(2\,\dot r\dot\phi+r\ddot\phi)\b
e_\phi+\ddot z\b e_z\,.
\ee
the smallness of the ratios $\beta_\perp^{}/\beta_\parallel^{}$ and $\beta_D^{}/\beta_\parallel^{}$, we arrive at an expression identical to Equation~(\ref{cur32}).

For $\beta_\perp^{}=0$ we have $\b a^2=a_0^2={\rm const}$. This is the case
when the particle moves strictly along the drift trajectory. For
$\beta_\perp^{}=\beta_D^{}$, the square of acceleration $\b a^2=2\,a_0^2\,(1-\cos\tau)$.
For time instants, when the acceleration becomes zero, the
particle velocity is parallel to the field line. In the limit of
$\beta_\perp^{}\gg \beta_D^{}$, the acceleration becomes independent of the time
and the radius of field curvature:

\be\label{cur34}
\b a^2=a_0^2\,\frac{\beta_\perp^2}{\beta_D^2}=
\left(\frac{c\beta_\perp^{}\omega_c}{\gamma}\right)^{\!2}\,.
\ee
It is well known (see e.g., \cite{Landau2}) that this
equation also describes the particle acceleration in the
homogeneous magnetic field. While for applicability of
Equation~(\ref{cur34}) the fulfillment of the condition $\beta_\perp^{}\gg \beta_D^{}$ is
sufficient, the relation between $\beta_\perp^{}$ and $\beta_\parallel^{}$ can be arbitrary.

For an ultrarelativistic particle, the acceleration and the
curvature radius of the trajectory are linked with a simple
relation $|\b a|=c^2/R_c$, which at $v=c$, is the definition of the curvature radius. 
Thus, the curvature radius of the trajectory is

\be\label{cur35}
R_c(t)=r_0\left(1-2\eta \cos\tau+\eta^2\right)^{\!-1/2}\, ,
\ee
where $\beta_\parallel$ is set to be equal to 1, given that $1-\beta_\parallel^{}\ll 1$. The
pitch angle $\alpha$ of the particle also depends on time:
\be\label{cur35a}
\alpha(t)=\beta_D^{}\left(1-2\eta \cos\tau+\eta^2\right)^{1/2},
\ee
and varies in the range
\be\label{cur35b}
|\beta_D^{}-\beta_\perp^{}|\le \alpha \le \beta_D^{}+\beta_\perp^{}\,.
\ee
Note that the angle between drift trajectory and particle velocity
is constant and equal to $\beta_{\perp}$.

\section{Intensity and energy spectrum of radiation}\label{sec:inten}

In this section we compute, within the framework of
classical electrodynamics, the intensity and the energy
spectrum of the radiation of a charged ultrarelativistic particle
with the perpendicular component of velocity $\beta_\perp\sim\beta_D$ when
both the curvature of the drift trajectory and the curvature of the
gyration are important. If the radius of the curvature of the
trajectory $R_c\sim r_0$, the characteristic energy of the emitted
photon is $\sim \hbar c\gamma^3/r_0$. The requirement of smallness of this
energy compared to the energy of the particle $mc^2\gamma$ (the
condition of applicability of classical electrodynamics) implies
$\hbar\gamma^2\ll mcr_0$. We assume that the perpendicular (relative to the
drift trajectory) momentum is $p_\perp^{}\ll p_\parallel^{}$, but, at the same time,
$p_\perp^{}\gg mc$. Using Equation (\ref{cur28}), the latter inequality can be
presented in the form $c\gamma^2\gg \omega_c r_0$. This gives the following
constraint on the Lorentz factor:

\be\label{cond}
\omega_c r_0/c \ll \gamma^2 \ll mcr_0/\hbar\,.
\ee
The condition of the smallness of the lower limit compared
to the upper limit, implies $\hbar\omega_c\ll mc^2$ (i.e., the magnetic field
should not exceed the critical field,
$B_{cr}=2m^2c^3/3e\hbar\approx 2.94\cdot 10^{13}$~G).

The energy radiated away by the particle during a unite time
interval is \citep{Landau2}
\be\label{cur36}
I=-\frac{dE}{dt}=\frac{2e^2\gamma^6}{3c^3}\,(\b a^2-(\b v\times \b a)^2/c^2 )
=\frac{2e^2c}{3}\,\frac{\gamma^4}{R_c^2}\,.
\ee
Substituting $R_c$ from Equation (\ref{cur35}), we obtain
\be\label{cur37}
I=\frac{2e^2c}{3}\,\frac{\gamma^4}{r_0^2} \left(1-2\eta\cos\tau+\eta^2\right).
\ee
The intensity averaged over time is
\be\label{cur38}
\langle I\rangle=\frac{2e^2c}{3}\,\frac{\gamma^4}{r_0^2} \left(1+\eta^2\right).
\ee
This formula can be written in as $\langle I\rangle=I_{\rm
curv}(1+\eta^2)$, where
$I_{\rm curv}$ is defined in Equation (\ref{cur8}). Thus, for a given Lorentz
factor the particle produces radiation with minimum intensity
during the curvature radiation regime.

In the considered scenario, we deal with three characteristic
times: the time of energy losses (cooling time),
$t_{\rm cool}\sim mcr_0^2/(e^2\gamma^3)$, the period of oscillations, $T\sim \gamma/\omega_c$, and
the characteristic time of formation of radiation, $\Delta t\sim r_0/(c\gamma)$.
It is easy to show, using Equation (\ref{cond}), that $\Delta t\ll t_{\rm cool}$. Note
that this condition is always satisfied in the framework of
classical electrodynamics. The ratio $\Delta t/T\sim \omega_c r_0/(c\gamma^2)$ also is
small compared to 1, as it follows from Equation (\ref{cond}). Finally,
for the ratio $T/t_{\rm cool} \sim e^2\gamma^4/(mc\omega_cr_0^2)$, using the lower and
upper limits of the Lorentz factor from Equation (\ref{cond}), we
obtain

\be\label{cond1}
\alpha_f^{}\,\frac{\hbar\omega_c}{mc^2}\ll \frac{T}{t_{\rm cool}} \ll
 \alpha_f^{}\,\frac{mc^2}{\hbar\omega_c}\,,
\ee
where $\alpha_f^{}=e^2/\hbar c\approx 1/137$ is the fine-structure constant.

The smallness of $\Delta t$ significantly simplifies the calculations
of the radiation spectrum. This allows us to ignore the changes
of the particle energy and the curvature radius with time, and
perform calculations at fixed prompt values for these
parameters: $E(t)=mc^2 \gamma(t)$ and $R_c(t)$. Then the spectral flux
density of the magnetic bremsstrahlung integrated over all the
emission angles is described by the well-known expression for
the synchrotron regime of radiation (see, e.g. \cite{Schwinger1949};
\cite{Bayer}):

\be\label{cur39}
P(\omega,t)=\frac{\sqrt{3}\,e^2}{2\,\pi}\,\frac{\gamma}{R_c}
\,F\!\left(\frac{\omega}{\Omega_*}\right).
\ee
Here
\be\label{cur40}
\Omega_*=\frac{3c\gamma^3}{2R_c}\,,
\ee
and
\be\label{cur41}
F(x)=x\int_x^\infty\! K_{5/3}(x')\,dx'\,,
\ee
where $K_{5/3}$ is a modified Bessel function. The particle energy
loss rate is determined to be $dE/dt=-\int_0^\infty\!P(\omega,t)\,d\omega$; the
calculation of this integral results, as expected, in Equation~(\ref{cur36}).

Remarkably, in all cases under consideration, the shape of
the energy spectrum is defined by the same function $F(x)$. The
relevant parameters only change the position of the maximum
and the intensity. However, the function $F(x)$ should be
replaced by its quantum analogue (see Equation (\ref{eq:fq})) if the
parameter $\chi=B \epsilon \sin\alpha/B_{cr} \geq 1$, where $\epsilon$ is the photon energy
in units of $m_{\rm e} c^2$, and $\alpha$ is the angle between the photon and the
magnetic field \citep{Landau4}. Note that at such
conditions the energy of the produced photon is close to the
energy of the radiating electron. The electron-positron pair
production by a gamma-ray photon in the strong magnetic field
occurs when $\chi\gtrsim 1$. In the curved magnetic field the angle
between the photon and the magnetic field could become
sufficiently large for production of electron-positron pairs. This
could lead to the development of an electromagnetic cascade,
provided that the optical depth is large. The cascade results in
the formation of radiation, the spectrum of which is
considerably different from that of the initial (synchrotron)
radiation.

The limits of the applicability of Equation (\ref{cur39}), which
describes the energy spectra of the synchrotron and curvature
radiation components, are determined by the approach
proposed by \cite{Schwinger1949}. The radiation of the ultrarelativistic
particle is concentrated in a narrow cone with the
opening angle $\sim 1/\gamma$, and therefore is collected when the angle
between the velocity and the direction of the observation is of
the order of the same as $\sim 1/\gamma$. The method of \cite{Schwinger1949} 
is based on the expansion of the trajectory in a small
time interval. During this time interval the entire observable
radiation should be collected. This means that the approach
works if the particle velocity changes direction at an angle
larger than $1/\gamma$ during the time that the expansion of the
trajectory is valid. The analysis of the local trajectory given by
Equations~(\ref{cur29a})–(\ref{cur29c}) gives the following limits of applicability
in the curved magnetic field:

\begin{eqnarray}\label{eq:applim}
\begin{aligned}
\beta_{\perp}&\gg \frac{1}{\gamma},\quad \text{if}\quad 
\beta_{\perp}\gtrsim \beta_D,\\
\sqrt{\frac{\beta_D^3}{\beta_D^3+\beta_{\perp}}}&\gg \frac{1}{\gamma}, \quad
\text{if} \quad \beta_{\perp}\lesssim\beta_D.
\end{aligned}
\end{eqnarray}
The first condition corresponds to the case of synchrotron
radiation and states that the perpendicular motion should be
relativistic as is expected from the consideration of radiation in
the homogeneous magnetic field. The second condition
corresponds to the situation of the synchro-curvature radiation
of a particle with a small perpendicular momentum. Small
gyrations around the drift trajectory do not influence the
applicability of Equation~(\ref{cur39}). In the limit $\beta_{\perp}=0$, this
condition simply implies that the motion should be relativistic.

Thus if the requirements of Equations~(\ref{eq:applim}) are fulfilled for
the synchrotron and curvature radiation regimes, the shape of
the prompt energy spectrum remains the same. The distinct
feature of the synchro-curvature regime is the significant time
variability of the curvature $R_c$. We assume that the condition
$T\ll t_{\rm cool}$ is satisfied (note that this condition does not
contradict Equation (\ref{cond1})). Then, since the change of $\gamma$ during
the period $T$ is small, we can introduce the parameters averaged
over time. This was already done at the changeover from
Equation (\ref{cur37}) to Equation (\ref{cur38}), assuming that during period $T$
the Lorentz factor $\gamma = {\rm const}$.

In accordance with Equation (\ref{cur35}),  
\be\label{cur42}
R_c=r_0/q(\eta,\tau)\,,\;\quad q(\eta,\tau)=(1-2\eta\cos\tau+\eta^2)^{1/2}\, . 
\ee
Here a new parameter $q$ determines the difference between
the curvature of the particle trajectory and the curvature of the
magnetic field line, and thus characterizes the deviation of the
particle radiation from the curvature radiation.

Now the averaged spectral flux density can be presented in
the form
\be\label{cur43}
\langle P(\omega,t)\rangle=\frac{\sqrt{3}\,e^2}{2\,\pi}\,\frac{\gamma}{r_0}
\,G\!\left(\frac{\omega}{\omega_*}\right) , 
\ee
where
\be\label{cur44}
G\!\left(\frac{\omega}{\omega_*}\right)=\frac{1}{\pi}\int_0^\pi\!
q(\eta,\tau)\,F\!\left(\frac{\omega}{\omega_* q(\eta,\tau)}\right) d\tau\,,
\ee
and $\omega_*=3c\gamma^3/(2r_0)$. $G$ is a function of two variables:
$x=\omega/\omega_*$ and $\eta=\beta_\perp^{}/\beta_D^{}$. If in Equation (\ref{cur44}) we represent
the function $F(x)$ in the form of Equation (\ref{cur41}), and change the
order of integration, the integral over $d\tau$ can be calculated
analytically. This allows us to express the function $G(x)$ in the
form of a single integral:
\be \label{eq:59}
G(x)=x\int_{x/(1+\eta)}^\infty\! K_{5/3}(x')\,\Psi(x/x')\,dx'\,,
\ee
where
\be
\Psi\!\left(\frac{x}{x'}\right)=\left\{
\begin{array}{ll}
\frac{1}{\pi}\arccos\frac{(x/x')^2-1-\eta^2}{2\eta}\,, & \frac{x}{x'}\ge
|1-\eta|\,,\\
1\,, & \frac{x}{x'}< |1-\eta|\,.
\end{array}
\right.
\ee

\begin{figure}
\begin{center}
\includegraphics[width=0.47\textwidth,angle=0]{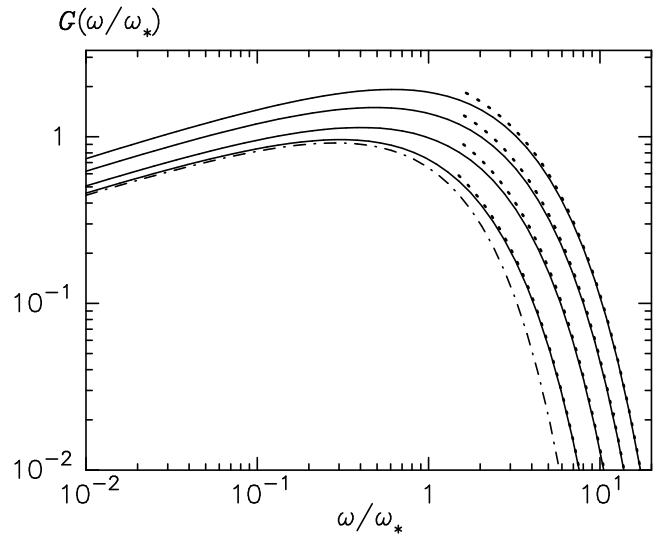}
\caption{\small 
Spectral flux density of the synchro-curvature radiation averaged
over time. The solid lines correspond to numerical calculations performed for
the values of the ratio $\eta=\beta_\perp^{}/\beta_D^{}$= = 0, 0.5, 1, 1.5 and 2 (from top to bottom,
respectively). The dotted lines are calculated using the asymptotic presentation
of the function $G(x)$ given by Equation (\ref{asymp}). The dot-dashed line corresponds
to the curvature radiation ($\eta=0$) when the particle moves along the drift
trajectory.
\label{f1}}
\end{center}
\end{figure}

The function $G$ for different values of η is shown in Figure~\ref{f1}.
At $\eta=0$ (the dot–dashed line) we have a nominal curvature
radiation. The curves correspond to the numerical integration of
Equation (\ref{cur44}), using for the function $F(x)$ the analytical
approximation given by Equation (\ref{eq:synchapp}). In two limits, $x \gg 1$
and $x \ll 1$, the function $F(x)$ can be expressed as

\begin{eqnarray}
F(x)\approx 2^{2/3}\Gamma\!
\left(\frac23\right)x^{1/3}\,,& \quad x\ll 1\,,\label{sy_as1}\\
F(x)\approx \sqrt{\frac{\pi x}{2}}\,e^{-x}\,,&
\quad x\gg 1\label{sy_as2}\,.
\end{eqnarray}
Therefore at $x\ll 1$, the function $G(x)$ differs from $F(x)$ only
by a factor that does not depend on $x$:
\be
G(x)=\frac{1}{\pi}\int_0^\pi\!(1-2\eta\cos\tau+\eta^2)^{1/3}\,d\tau\times
F(x)\,.
\ee
In the opposite limit, $x\gg 1$, the integral can be calculated
using the standard saddle point method. This gives the
following asymptotics:

\be\label{asymp}
G(x)\approx \frac{(1+\eta)^2}{2\sqrt{\eta}}\exp\!\left(-\frac{x}{1+\eta}\right)\,.
\ee
Formally, in this case the saddle point method can be applied if
$x\gg(1+\eta)^3/\eta$. However, the numerical calculations show
that in the interval $0.2\le \eta \le 3$ the simple analytical function
given by Equation (\ref{asymp}) already provides an accuracy better
than 10 \% at $x\approx 2$ (see Figure~\ref{f1}).

The spectrum described by Equations (\ref{cur43}) and (\ref{cur44}) for a
given magnetic field with curvature radius $r_0$ depends on the
value $\eta$ and Lorentz factor $\gamma$. In a course of motion and
radiation these parameters change, modifying the spectrum. In
numerical calculations described further it is more convenient
to use the $q$-parameter and Lorentz factors for the determination
of the spectrum. As it follows from Equation~(\ref{cur44}) the peak of
the distribution shifts with the change of parameters as $~q\gamma^3$.
This means that the peak of the spectra at a given energy
depends on the regime of the radiation determined by $q$.

The local form of $q$ given by Equation (\ref{cur42}) is useful for the
theoretical studies. In particular it is seen that $q$ oscillates
between $|1-\eta|<q<1+\eta$ with a frequency $\gamma$ times
smaller than the cyclotron frequency. With loss of energy the
oscillation becomes faster. For $\eta<1$, the $q$-parameter
oscillates around $1$, sweeping the bandwidth $2\eta$. At $\eta>1$ the
oscillations occurs around $\eta$ with an amplitude of $2$. Taking
into account Equation~(\ref{cur32}) the $q$-parameter can be written as
\begin{equation}\label{eq:qgen}
q=\frac{a}{a_0}=\frac{evB\sin\alpha}{mc\gamma}/\frac{v_{\parallel}^2}{r_0}\approx \frac{\sin\alpha}{\beta_D},
\end{equation}
where $\alpha$ is the pitch angle, $v$ is the particle velocity, and $v_{\parallel}$ is
the velocity along the magnetic field. The approximation sign
shows that we have used the condition $v\sim c$. Moreover, this
expression assumes that the conditions given by Equation~(\ref{cur30})
are fulfilled. This form is particularly convenient in numerical
calculations possessing all the properties of the local form, as
will be confirmed below. The value of the $q$-parameter allows
us to follow the evolution of the radiation regime. If radiation
proceeds in the curvature regime, the $q$-parameter is close to $1$
with small variations ($\eta\ll 1$) around it. When the amplitude of
oscillations is of the order of the average value of q, the
radiation proceeds in the synchro-curvature regime. When the
average value of $q$ is much larger than the amplitude ($\Delta q=2$)
of oscillations, the particle radiates in the synchrotron regime,
where the $q$-parameter could be very large. Correspondingly,
the radiation spectra are shifted to higher energies. An
important conclusion that follows from the consideration of
$q$-parameter is that the spectra produced in the curvature
radiation regime ($q\approx 1$) is shifted to the lower energies in
comparison with the spectra produced in the synchrotron and
synchro-curvature regimes.

For the derivation of the average spectral distribution we can
use also a different approach. Equation~(\ref{cur44}) can be presented in
the form ($\alpha\ll 1$)
\be\label{curE1}
G\!\left(\frac{\omega}{\omega_*}\right)=\frac{1}{\pi}\!\int_0^\pi\!\!d\tau
\!\int_0^\infty\!\!\! d\alpha\,\d(\alpha-\beta_D^{}q(\eta,\tau))
\frac{\alpha}{\beta_D^{}}\,
F\!\left(\frac{\omega\beta_D^{}}{\omega_* \alpha}\right).
\ee
After trivial integration over $d\alpha$, we obtain Equation~(\ref{cur44}). The
function
\be\label{curE2}
p(\alpha)=\frac{1}{\pi}\int_0^\pi\!d\tau\,
\d(\alpha-\beta_D^{}q(\eta,\tau))
\ee
describes the distribution of particles over pitch angles, thus
$p(\alpha)\,d\alpha$ is the probability of the pitch angle being between
$(\alpha,\,\alpha+d\alpha)$. Therefore Equation~(\ref{curE1}) can be written in the
form
\be\label{curE3}
G\!\left(\frac{\omega}{\omega_*}\right)=
\int_0^\infty\!\!\! d\alpha\,\frac{\alpha}{\beta_D^{}}\,p(\alpha) \,
F\!\left(\frac{\omega\beta_D^{}}{\omega_* \alpha}\right) .
\ee
As a result $G$ can be interpreted as the function $F$ averaged over
distribution of pitch angles. Note that in the derivation of
Equation~(\ref{curE3}) the form of the function $q$ is not specified, so
Equations~(\ref{curE3}) and (\ref{cur44}) are equivalent.

The determination of the spectral distribution of curvature
radiation through integration over the pitch angles has been
used by \cite{Epstein} (see Equation (20) in that paper).
However, one has to be cautious when applying this method.
The pitch angle of particles moving in the field with curved
lines depends on time (contrary to the case of motion in the
homogeneous magnetic field). Therefore the distribution $p(\alpha)$
cannot be adopted in an arbitrary manner, as was done by
\cite{Epstein}. This function should be derived from
Equation (\ref{curE2}) or obtained from the solution of relevant kinetic
equations. For example if $q(\eta,\tau)$ is given by Equation (\ref{cur42}),
we obtain the following expression:

\be\label{curE4}
p(\alpha)=\frac{2\,\alpha}
{\pi\sqrt{\big(\beta_D^2(\eta+1)^2-\alpha^2\big)
\big(\alpha^2-\beta_D^2(\eta-1)^2\big)}}
\ee
in the range of pitch angles $\beta_D^{}|\eta-1|\le \alpha \le
\beta_D^{}(\eta+1)$, and
$p(\alpha)=0$ outside that interval. Note that at $\eta\ne 1$, we do not
have particles with a null pitch angle. The integration of
Equation~(\ref{curE3}) over this distribution function gives the same
result as that obtained by integration over the time.

Let us consider now the case when the accelerated electrons
are not strictly unidirectional but are distributed within a
narrow cone with an opening angle $\Delta\psi\ll 1$ and axis along the
drift trajectory. We assume that all electrons have the same
energy. For the standard synchrotron radiation, a slight change
of the pitch angle does not affect the spectrum of radiation.
However, in the case of the synchro-curvature radiation even a
tiny spread of the angles relative to the drift trajectory may
result in a significant change of the spectrum. It is convenient to
express the angular distribution of particles through the
parameter $\eta$, which is linked to the the angle between the
velocity and the drift trajectory with a simple relation:
$\psi=\eta\beta_D^{}/\beta_\parallel^{}$. Note that at $\eta\lesssim 1$, the angle $\psi\ll 1$. For
calculations, the angular distribution of particles should be
specified. Below we consider a Gaussian type distribution:

\be\label{dis1}
g(\eta)\,d\eta \sim e^{-\eta^2/(2\zeta)}\, d\eta\,,
\ee
with the mean value of the square of the perpendicular velocity
\be\label{dis2}
\langle\beta_\perp^2\rangle= \zeta\,\beta_D^2\,.
\ee
For the determination of the energy spectrum we have to
average Equation~(\ref{cur44}) over the time $\tau$ and the parameter $\eta$,
i.e., calculate the integral

\be\label{cur45}
\langle G(x)\rangle=\int_0^\infty \! d\eta\,g(\eta)\int_0^\pi\!\frac{d\tau}{\pi}
q(\eta,\tau)\,F\!\left(\frac{x}{q(\eta,\tau)}\right).
\ee

\begin{figure}
\begin{center}
\includegraphics[width=0.47\textwidth,angle=0]{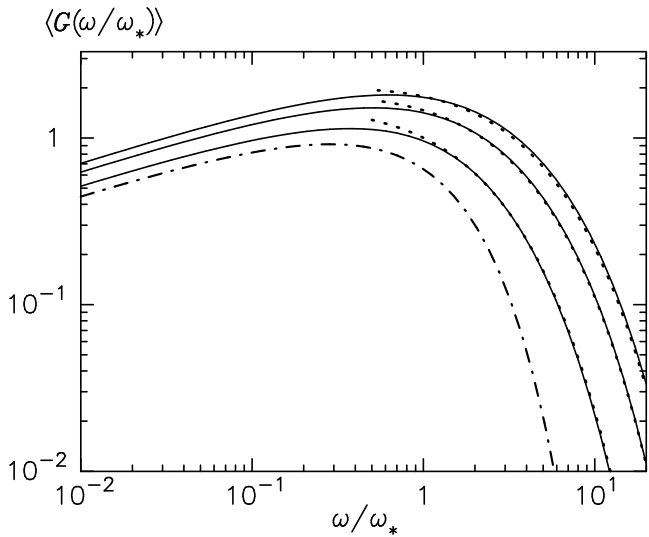}
\caption{\label{f2} \small 
Spectral flux density of the synchro-curvature radiation averaged
over the time, and the angle between the velocity and the drift trajectory,
assuming a Gaussian type distribution given by Equation~(\ref{dis1}). The solid lines
correspond to different values of $\zeta=\langle\beta_\perp^2\rangle/\beta_D^2$: 
1, 3, and 5 (from bottom to top, respectively). The dotted lines are calculated using the asymptotic
analytical presentation given by Equation~(\ref{dis3}). The dashed-dotted line
corresponds to the curvature radiation ($\zeta=0$).
}
\end{center}
\end{figure}

The energy spectra calculated for several values of $\zeta$ are
shown in Figure~\ref{f2}. The case of $\zeta=0$ corresponds to the
spectrum of the curvature radiation.

As above, using the saddle point method, we can find the
asymptotics for $\langle G\rangle$ at large $x$. Rather simple although
cumbersome computations, which we omit here, lead to the
following result

\be\label{dis3}
 \langle G(x)\rangle\approx\sqrt{\frac{\zeta}{3x^{1/3}}}
\left(x^{1/3}+\zeta^{-1/3} \right)^2\,e^{-u}\,,
\ee
where
\be\label{dis4}
 u=\frac32\,\zeta^{-1/3}x^{2/3}-\zeta^{-2/3}x^{1/3}+\frac1{3\zeta}\,.
\ee
This equation is derived under the following conditions: $x\gg \zeta$
and $x\gg 1/\zeta$. However, the comparison of Equation~(\ref{dis4}) with
accurate numerical calculations shows (see Figure~\ref{f2}) that for
$\zeta\sim 1$ this analytical presentation gives correct results at
$x\gtrsim 0.5$.

\begin{figure}
\begin{center}
\includegraphics[width=0.47\textwidth,angle=0]{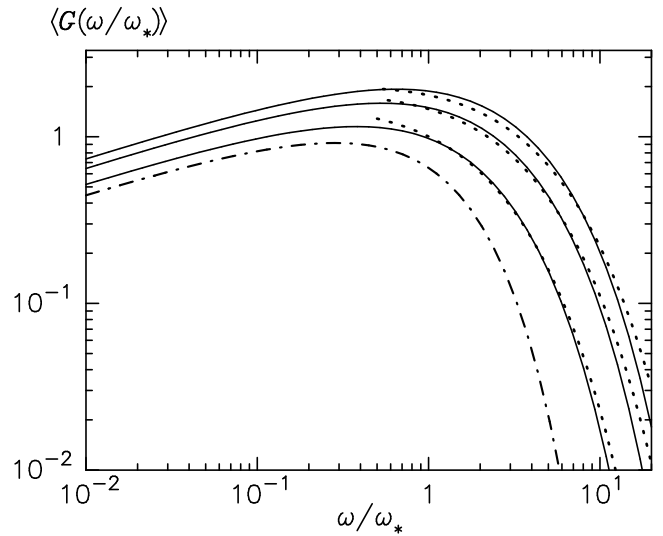}
\caption{\label{flat}\small 
Spectral flux density of the synchro-curvature radiation averaged
over the time and angle between the velocity and the drift trajectory, assuming
a uniform angular distribution of electrons within the opening angle of the
beam. The solid lines correspond to different values of the ratio 
$\zeta=\langle\beta_\perp^2\rangle/\beta_D^2$:
1, 3, and 5 (from bottom to top, respectively). The dotted lines are the same as
in Figure~\ref{f2}. They are calculated for the Gaussian type of angular distribution
using the asymptotic analytical presentation given by Equation~(\ref{dis3}). The
dashed-dotted line corresponds to the curvature radiation ($\zeta=0$).
}
\end{center}
\end{figure}

Although the assumed Gaussian type angular distribution of
electrons seems to be a quite reasonable and natural choice in
the considered scenario, it would be interesting to investigate
the dependence of the radiation spectrum on the specific
angular distribution of the electron beam. In particular, in
Figure~\ref{flat}, we show the radiation spectra calculated for the
uniform distribution of electrons within the fixed opening
angles of the beam. The solid curves in Figure~\ref{flat} are obtained
for the same values of the mean square of the perpendicular
velocity as in Figure~\ref{f2}. In Figure~\ref{flat} we show the asymptotic
solutions given by Equation~(\ref{dis3}) for the spectra calculated for
the Gaussian type angular distribution (the same dotted curves
from Figure~\ref{f2}). Figure~\ref{flat} demonstrates that the value of the
mean square of the perpendicular component of velocity
describes the energy spectrum of radiation quite well,
independent of details of the specific small pitch-angle
distribution of electrons.

Note that the spectra of the synchro-curvature radiation do
not contain the characteristic exponential term $e^{-x}$, as is the
case of the synchrotron or curvature radiation, but show a
harder behavior. This can be seen from the asymptotics in
Equation~(\ref{dis4}). To demonstrate the tendency of the steepening
of the energy spectrum beyond the maximum at $x \sim 1$, in
Table~\ref{tab:1} we show the evolution of the local slope of the spectral
flux density (the spectral index $s$), which is determined to be
\be
s=-d(\log \langle G\rangle)/d(\log\omega)\,.
\ee
It is seen that whereas in the case of curvature radiation ($\zeta=0$)
the spectrum becomes very steep at $x\gtrsim 3$, in the synchro-curvature
radiation regime with $\zeta=2$ or $\zeta =3$ a relatively
hard spectrum can extend up to $x\sim 10$.

\begin{table}
{\caption{\label{tab:1} The spectral index $s$ at different points $x=\omega/\omega_*$ of the spectrum $\langle G(x)\rangle$ given by Eq.~(\ref{dis3}) for various values of $\zeta=\langle\beta_\perp^2\rangle/\beta_D^2$.}}
\begin{tabular}{c||cccccc}
\hline
\hline
&2.0 & 3.0 &5.0 & 10. & 20. &30.\\
\hline
0 &1.65 &2.62 &4.59 & 9.56 & 19.5 & 29.5\\
\hspace{3pt}0.5\hspace{3pt}&\hspace{3pt}1.17\hspace{3pt}&\hspace{3pt}1.67\hspace{3pt} &\hspace{3pt} 2.56\hspace{3pt} &\hspace{3pt} 4.45\hspace{3pt} & \hspace{3pt}7.56\hspace{3pt} &\hspace{3pt}10.2\; \hspace{3pt}\\
1.0 &0.96 &1.37& 2.10 & 3.63 & 6.14 & 8.28\\
2.0 & 0.75 & 1.08 &1.67 & 2.91 & 4.93 & 6.65\\
3.0 & 0.64&0.93 &1.45 & 2.54& 4.31& 5.82\\
\hline
\end{tabular}
\put(-217,30.2){\line(5,-2){24}}
\put(-215,20){${}^\zeta$}
\put(-200,23.5){${}^x$}
\end{table}

\section{Numerical implementation}\label{sec:num}
The analytical approach described above allows us to study
the local properties of the particle trajectory and its radiation in
the curved magnetic field. To solve the problem in the general
case, namely, to take into account the energy losses of the
particle during its propagation, we performed numerical
integration of the equations of motion. The radiation properties
have been studied for the dipole magnetic field. This is a good
approximation for the case of a strong magnetic field in the
vicinity of compact astrophysical objects. As it is shown in
\cite{Deutsch1955}, the instantaneous configuration of the
magnetic field in the vicinity of a rotating star appears as a
stationary magnetic dipole. Furthermore, in Section~\ref{sec:elec}, we use
the approximation given by \cite{Deutsch1955} for the electric
field in the vicinity of the star. The dipole magnetic field has
two distinct features. The first is the fast decrease of its strength
with the distance, $B(r) \propto 1/r^3$, which has a strong impact on
the radiation intensity. The second is the significant variation of
the curvature of the field lines, with a change of the polar angle
$\theta$ from the dipole axis $\propto \sin\theta/r$. Therefore the radiation spectra
in the vicinity of the pole and the equator have different
characters. 

For the calculation of the exact trajectory of a particle in the
curved magnetic field, we have used the equations of motion in
the ultrarelativistic limit. In this limit the velocity has the
constant value of the speed of light and changes only its
direction. The radiation reaction force is directed opposite to
the velocity. Thus the equation of motion takes the form
\begin{equation}\label{eq:moteq}
mc\left(\b\beta\frac{d\gamma}{dt}+\gamma\frac{d\b \beta}{dt}\right)=e(\b \beta\times \b B)-|\b f|\b \beta,
\end{equation}
where $\b \beta$ is the velocity in units of $c$ with $|\b \beta|\approx 1$, $\b B$, B is the
magnetic field, and $\b f$ is the radiation reaction force \citep{Landau2}. 
Here, the left side of the equation is expanded
for the convenience of further transformations. Taking the
scalar product of Equation~(\ref{eq:moteq}) with velocity  $\b \beta$ and using the
fact that $\b \beta^2=1$ and, therefore, $\b\beta\frac{d\b\beta}{dt}=0$, we obtain the
following differential equation for the Lorentz factor
\begin{equation}\label{eq:dgamma}
\frac{d\gamma}{dt}=-\frac{|\b f|}{mc}.
\end{equation}
Substitution of this equality back to Equation~(\ref{eq:moteq}) gives us
\begin{equation}\label{eq:ofmotion}
\frac{d\b \beta}{dt}=\frac{e}{mc\gamma}(\b \beta\times \b B),
\end{equation}
which has the same form as if we considered the motion
without energy losses. However, we now have two equations
where the Lorentz factor is variable: Equations~(\ref{eq:dgamma}) and (\ref{eq:ofmotion}).
For the sake of convenience of the numerical treatment and
comprehension of the structure of the system of equations,
these equations (plus the equation for coordinates) can be
written in the dimensionless form (see also Appendix~\ref{sec:accel}):
\begin{align}
\frac{d\b r'}{d \tau}&=u_1 \b \beta\,, \\
\frac{d\b \beta}{d\tau}&=\xi(\b \beta \times\b b)\,, \\
\frac{d\xi}{d\tau}&=u_2(\b \beta \times\b b)^2\,\label{eq:LossEn}.
\end{align}
The system of equations depends on two dimensionless parameters
\begin{equation}
u_1=\frac{mc^2\gamma_0}{eB_0R_0} \quad  {\rm and} \quad
u_2=\frac{2}{3}\frac{e^3B_0}{m^2c^4}\gamma_0^2,
\end{equation}
where $\gamma_0$ is the initial Lorentz factor, $R_0$ is the characteristic
distance to the radiating region from the dipole (the distance to
the outer gap or polar cap), $B_0=B_* \left(R_*/R_0\right)^3$ is the
characteristic magnetic field at this distance with $R_*$ and $B_*$
being the star radius and the magnetic field at the magnetic pole
of the star, $m$ is a particle mass, and $c$ is the speed of light. Here
we have introduced the following dimensionless variables: $\b r'$ is
the coordinate in the units of $R_0$, $\b \beta$ is the velocity of the particle
in the units of $c$, $\xi=\gamma_0/\gamma$ is the ratio of the initial and current
value of the Lorentz factor, $\tau=t\,eB_0/mc\gamma_0$ is the characteristic
time in units of the initial gyration period, and $\b b$ is the
dimensionless dipole magnetic field that is expressed as
\begin{equation}
\b b=\frac{3\b n(\b n \b \mu)-\b \mu}{2r'^3},
\end{equation}
where $\b \mu$ is the unit vector in the direction of the dipole axis,
and $\b n=\b r'/r'$ is the unit vector in the direction of the particle
position.

It should be noted that, depending on the specific conditions
characterizing an astrophysical source, the parameters $u_1$ and $u_2$
may differ from each other by many orders of magnitude. For
instance, for typical parameters of the polar cap model of the
pulsar magnetosphere $B_0=10^{12}$ G, $R_0=10^6$ cm, and
$\gamma_0=10^8$, we have $u_1\approx 1.7\times 10^{-7}$ and $u_2\approx 1.1\times 10^{12}$.
Thus, we deal with a non-stiff problem. The implicit
Rosenbrock method has been used for integration of this
system of differential equations.

The calculations of the trajectory were performed for
different initial conditions. The initial position is determined
by the radius $R_0$ and the polar angle $\theta_0$ relative to the magnetic
dipole axis. The radiation spectrum was calculated for different
initial pitch angles $\alpha_0$ and the initial Lorentz factors $\gamma_0$. The
information regarding the particle trajectory and its energy
allows us to derive the $q$-parameter from Equation~(\ref{eq:qgen}), and
then calculate the radiation spectrum at any moment of time
using Equation~(\ref{cur44}). The {\it cumulative spectrum} (i.e., the
spectrum integrated over the time), can be calculated for any
given initial parameters.

Finally, in the case of a very strong magnetic field and/or
very large Lorentz factor, the particle may radiate in the
quantum regime. More specifically, when the parameter
$\chi=B\gamma \sin\alpha/B_{cr} \geq 1$, Equation~(\ref{eq:LossEn}) should be replaced by
its quantum analogue (see Equation~(\ref{eq:QLossApp}))
\begin{equation}
\frac{d\xi}{d\tau}=\frac{u_2(\b \beta \times\b b)^2}{\left(1+u_3
((\b \beta \times\b b)/\xi)^{2/3} \right)^2 },
\end{equation}
where $u_3=1.07\times 10^{-9}(B_0\gamma_0)^{2/3}$.

\section{Astrophysical implications}\label{sec:implic}
In this section we explore some possible realizations of the
synchrotron and curvature regimes, as well as transitions
between these two regimes of radiation (the synchro-curvature
regime) in the context of two specific astrophysical scenarios.
Namely, we discuss the radiation of electrons and positrons in
the pulsar magnetosphere for the {\it outer gap} and {\it polar cap}
models.

Before proceeding, we first make some estimations of the
basic parameters in the dipole magnetic field. The radius of
light cylinder is ${R_{lc}\approx1.57\cdot10^{8}P_{cr}}$~cm where ${P_{cr}=P/33}$~ms
and $P$ is the period of pulsar. The strength of magnetic field is
${B\approx1.28\cdot10^{5}B_{12}P^{-3}_{cr}R^{-3}}$~G, where $B_{12}=B/10^{12}G$ and $R=r/R_{lc}$. 
The drift velocity is $\beta_{D}\approx1.26\cdot10^{-3}\gamma_{7}R^2B^{-1}_{12}P^2_{cr}\sin\theta$,
where ${\gamma_7=\gamma/10^{7}}$, and $\theta$ is the polar angle
from dipole axis. Here we use the formula $\beta_D=\gamma mc^2/eBr_0$,
where $r_0$ is the curvature of the magnetic field lines and $\beta_{\parallel}=1$
(compare with Equation~(\ref{cur28})). Then according to Equation~(\ref{eq:qgen}) 
${q\approx0.79\cdot10^{3}\alpha\gamma^{-1}_{7}R^{-2}B_{12}P^{-2}_{cr}\sin^{-1}\theta}$,
where $\alpha$ is
the pitch angle. The characteristic energy of the curvature
radiation is ${\epsilon_{curv}\approx2.8\cdot10^{8}\gamma^3_7R^{-1}P^{-1}_{cr}\sin\theta}$~eV.
Thus, an electron with $q=\alpha/\beta_D$, which determines the regime of the
radiation emits photons with characteristic frequency
$\epsilon_{0}=q\epsilon_{curv}$ with the intensity $I=q^2I_{curv}$, where $I_{curv}$ is the
intensity of the curvature radiation.
 
\vspace{1cm}

\subsection{Outer Gap}
\begin{figure*}
\begin{center}
\mbox{\includegraphics[width=0.5\textwidth,angle=0]{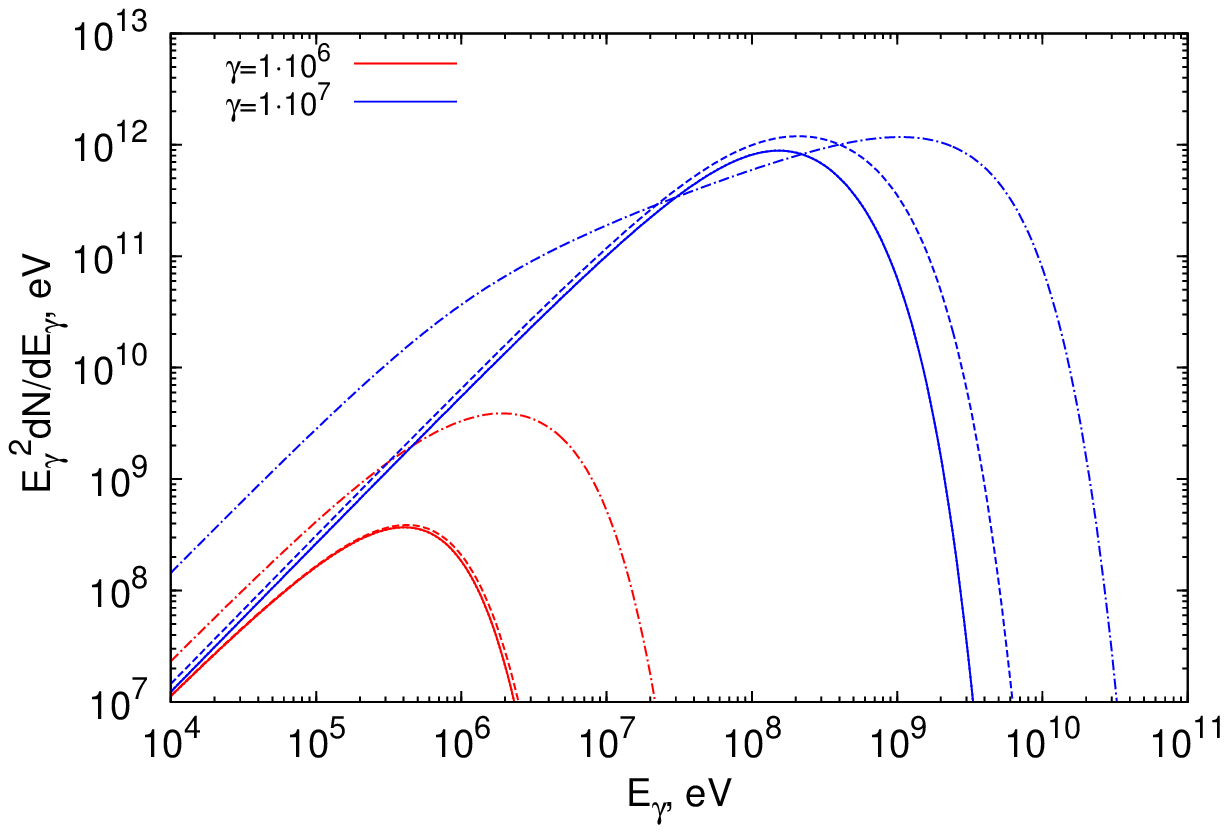}
\includegraphics[width=0.5\textwidth,angle=0]{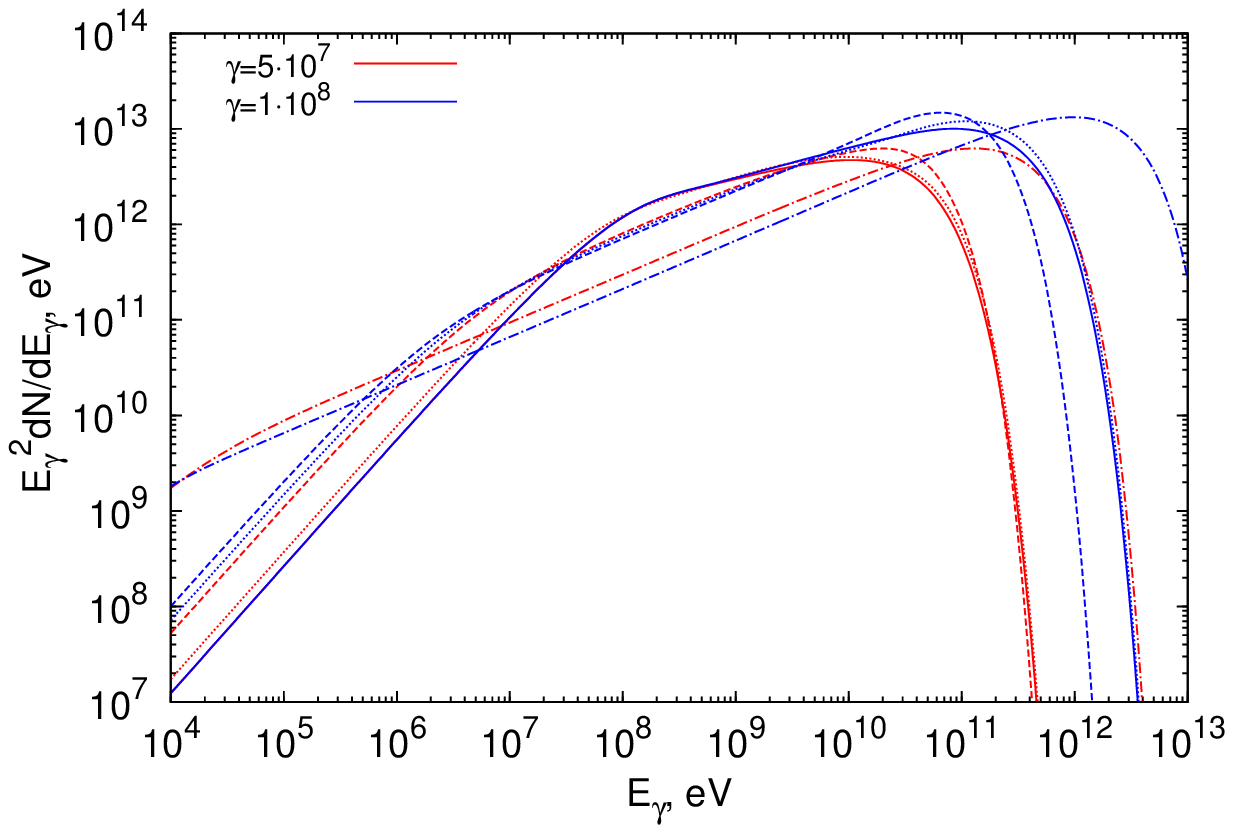}}
\caption{\label{fig:OGrad} 
 Cumulative (integrated along trajectory) radiation spectra of electrons calculated for the outer gap model in the pulsar magnetosphere. The curves are
 obtained for the initial Lorentz factors of electrons $\gamma=10^6$ and $10^7$ (left panel), and $\gamma=5\times 10^7$ and $10^8$ (right panel). The spectra calculated in the approximation
 of the motion strictly along magnetic field lines are shown by solid lines. Other spectra are calculated using the exact trajectories of electrons with different initial
 directions relative to the magnetic field lines: along the drift trajectory (dotted lines) (not seen on the left panel because of the coincidence with solid lines), along the
 magnetic field line (dashed lines), and at pitch angle $\alpha=10\beta_D$ in the meridional plane opposite to the normal vector of magnetic field lines (dashed-dotted lines).
}
\end{center}
\end{figure*}

\begin{figure}
\begin{center}$
\begin{array}{cc}
\includegraphics[width=0.25\textwidth]{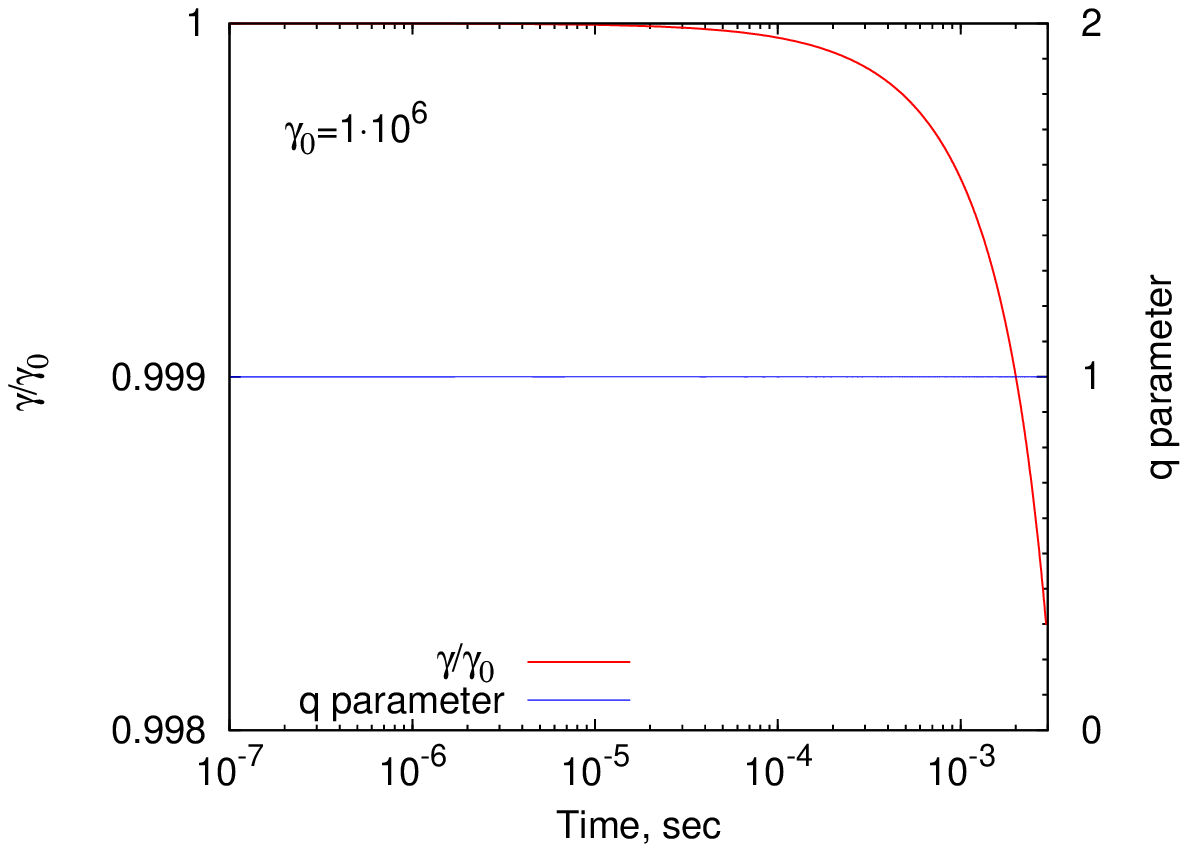} &
\includegraphics[width=0.25\textwidth]{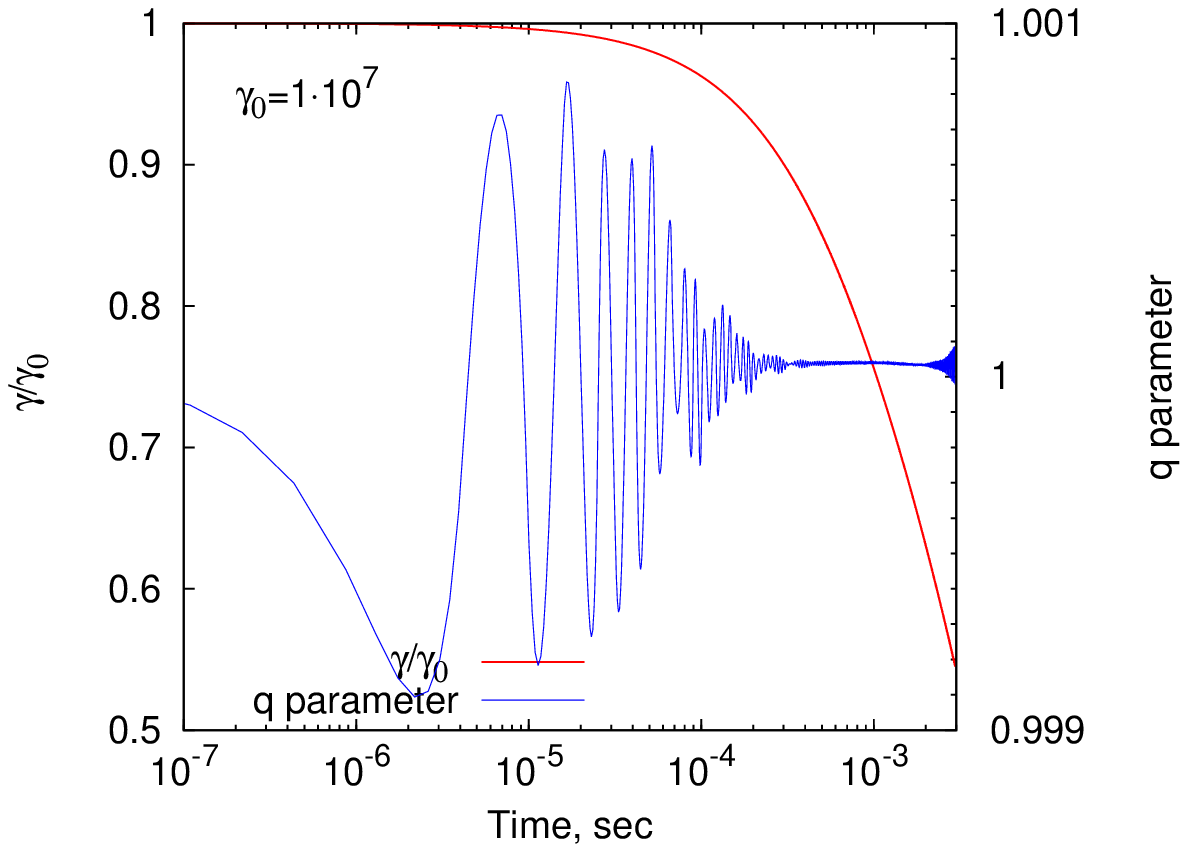} \\
\includegraphics[width=0.25\textwidth]{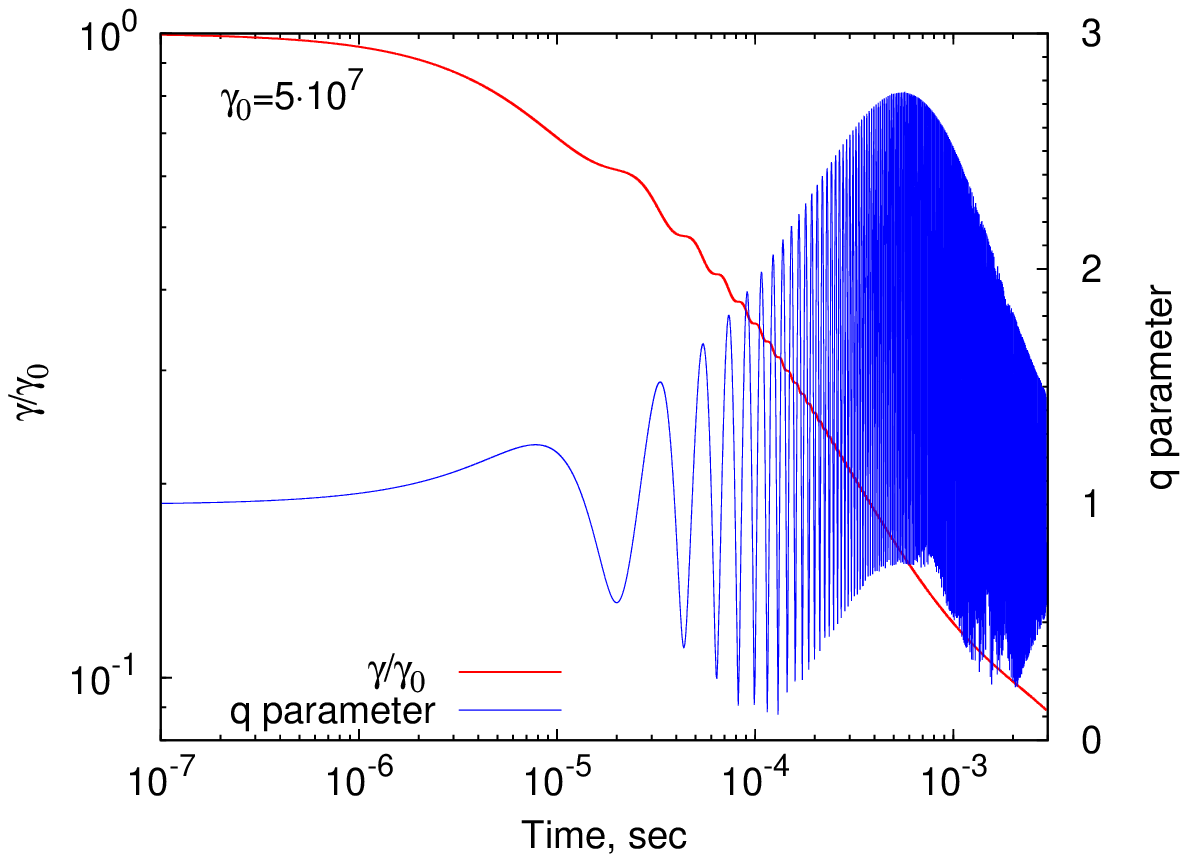} &
\includegraphics[width=0.25\textwidth]{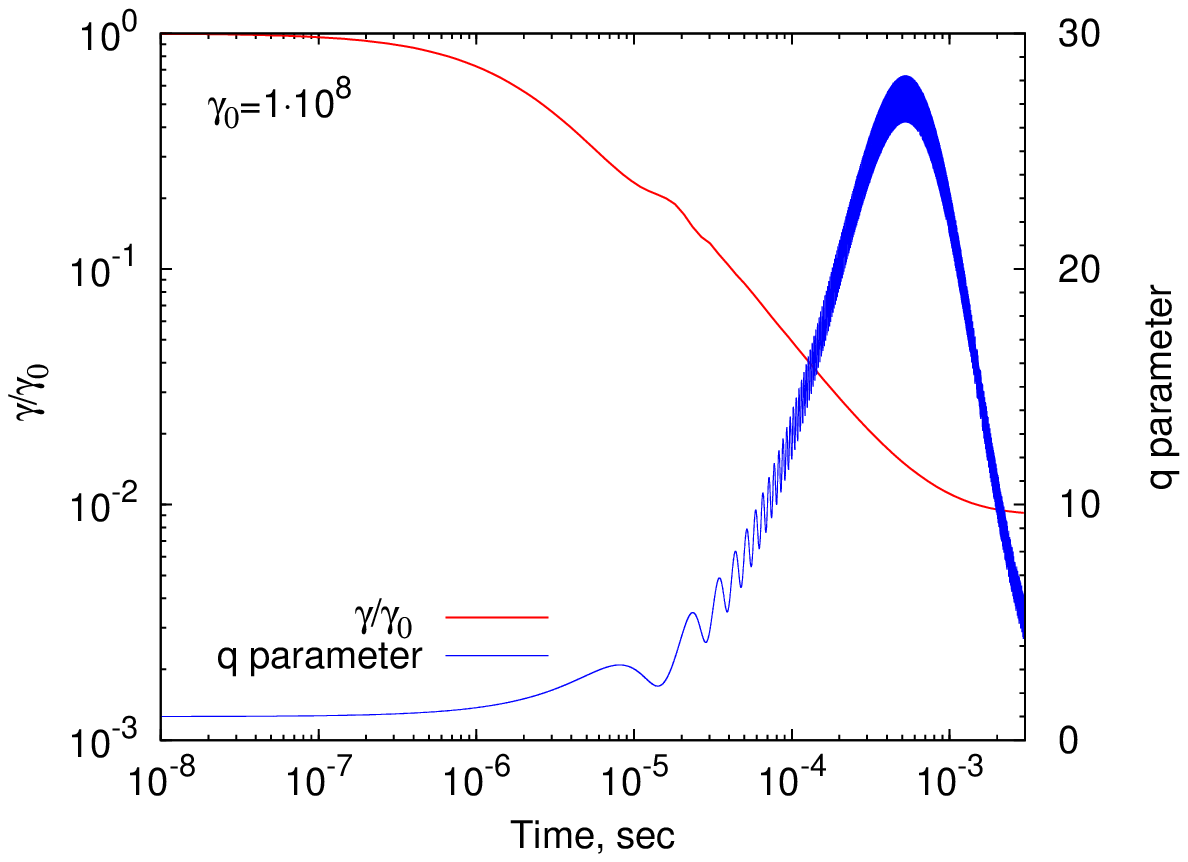}
\end{array}$
\end{center}
\caption{\label{fig:OGradComD}
 Time evolution of the q-parameter and the electron Lorentz factor in the outer gap model 
 (complementary to Figure~\ref{fig:OGrad}). The four panels correspond to the
 initial Lorentz factor of electrons $\gamma=10^6, 10^7, 5\times 10^7, 10^8$ and their initial direction 
 along the drift trajectory.
}
\end{figure}

\begin{figure}
\begin{center}$
\begin{array}{cc}
\includegraphics[width=0.25\textwidth]{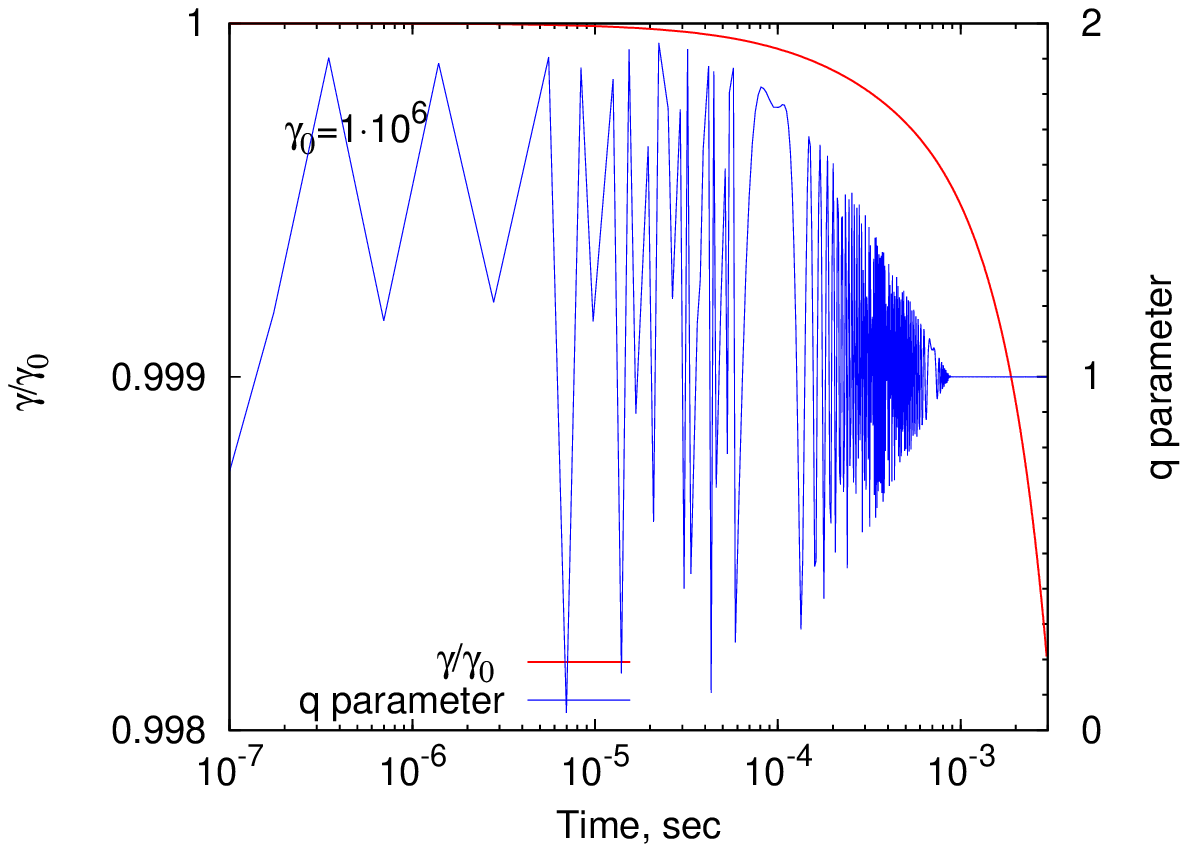} &
\includegraphics[width=0.25\textwidth]{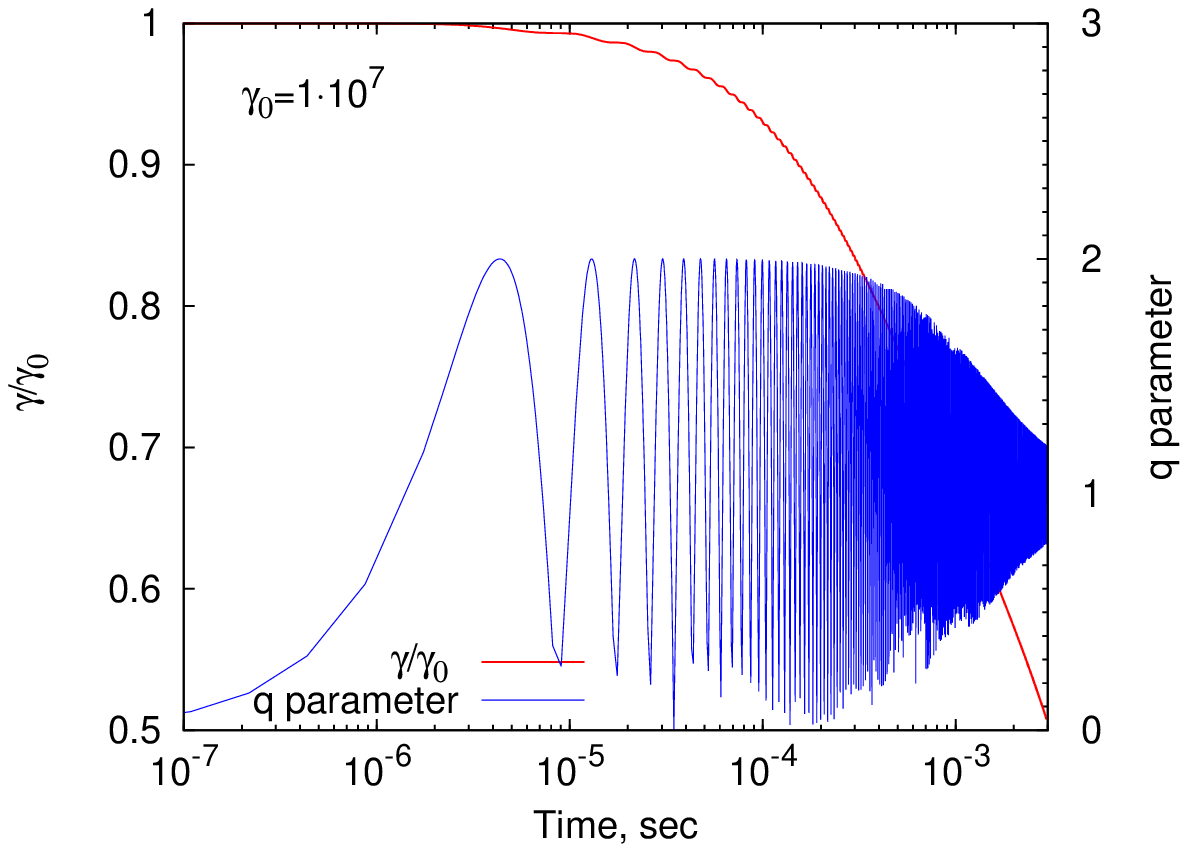} \\
\includegraphics[width=0.25\textwidth]{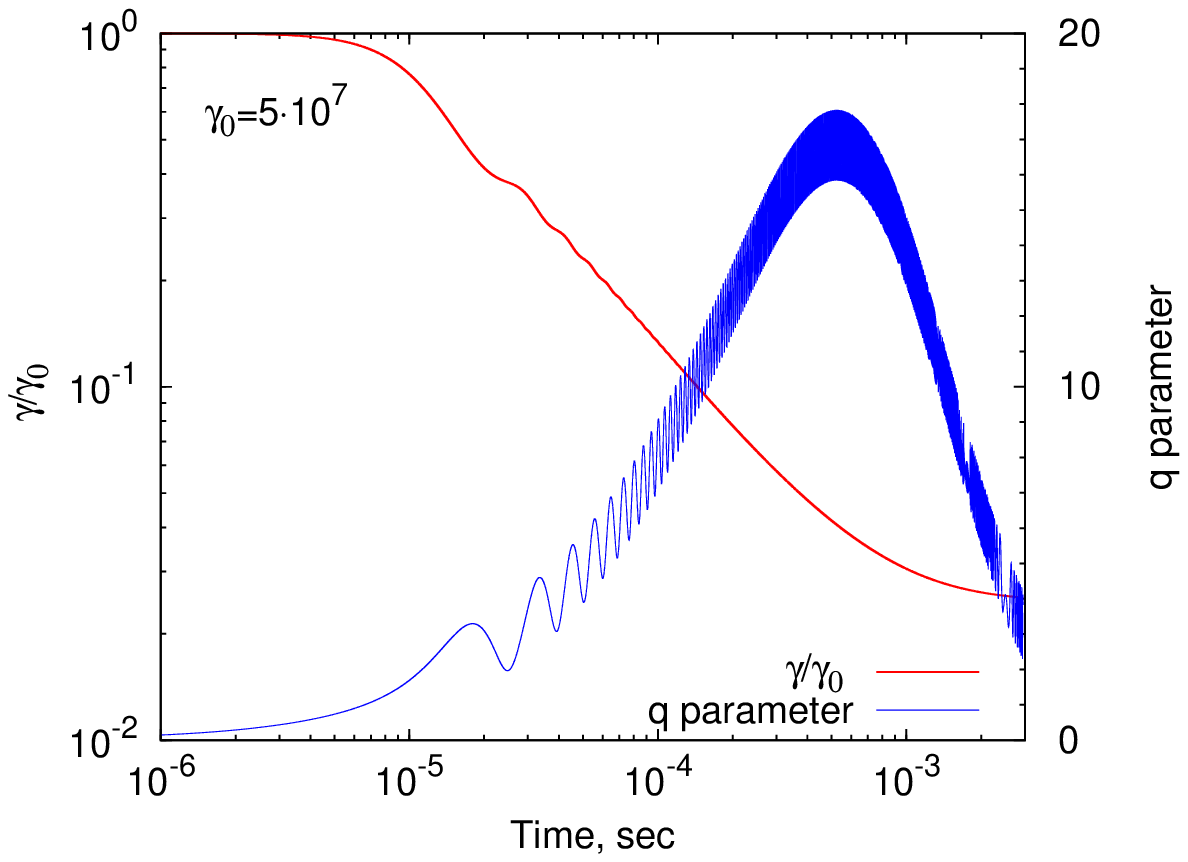} &
\includegraphics[width=0.25\textwidth]{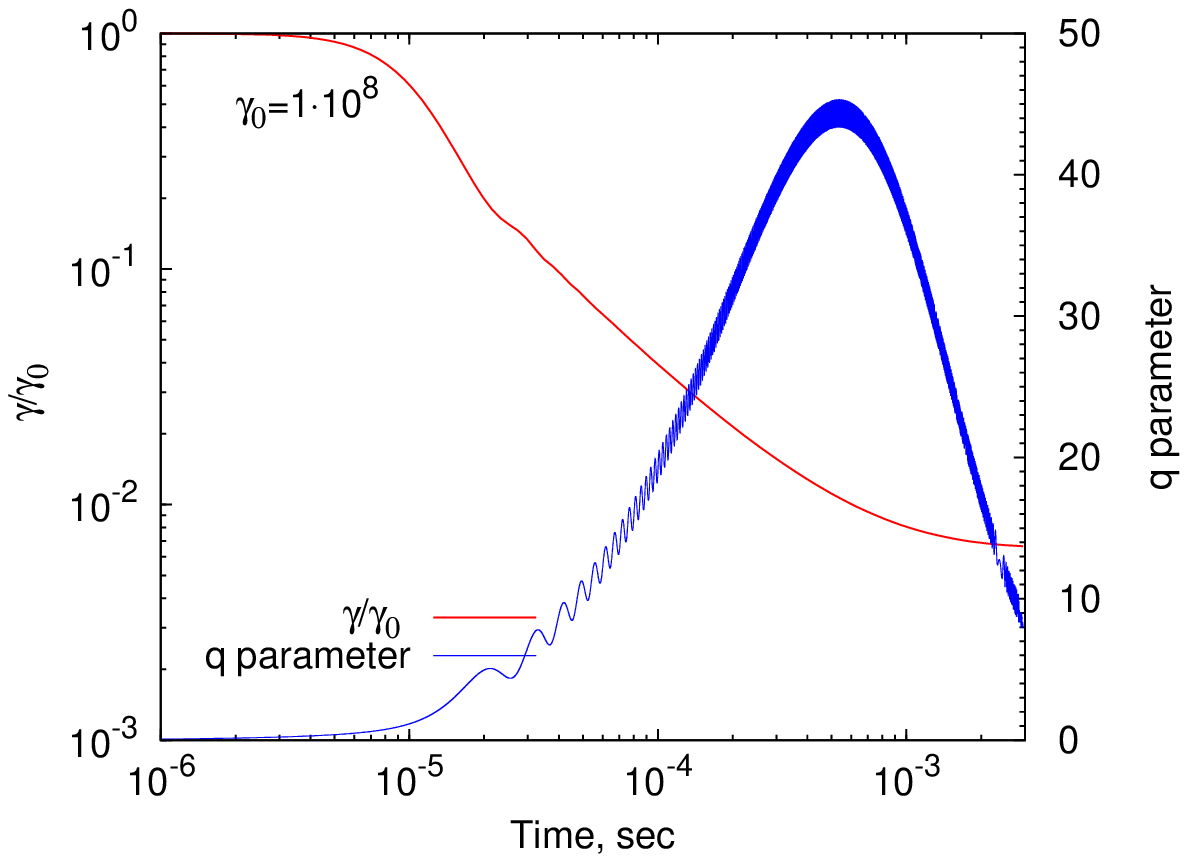}
\end{array}$
\end{center}
\caption{\label{fig:OGradComB}
Same as in Figure~\ref{fig:OGradComD}, but for the initial direction of electrons along the magnetic field line.
}
\end{figure}

\begin{figure}
\begin{center}$
\begin{array}{cc}
\includegraphics[width=0.25\textwidth]{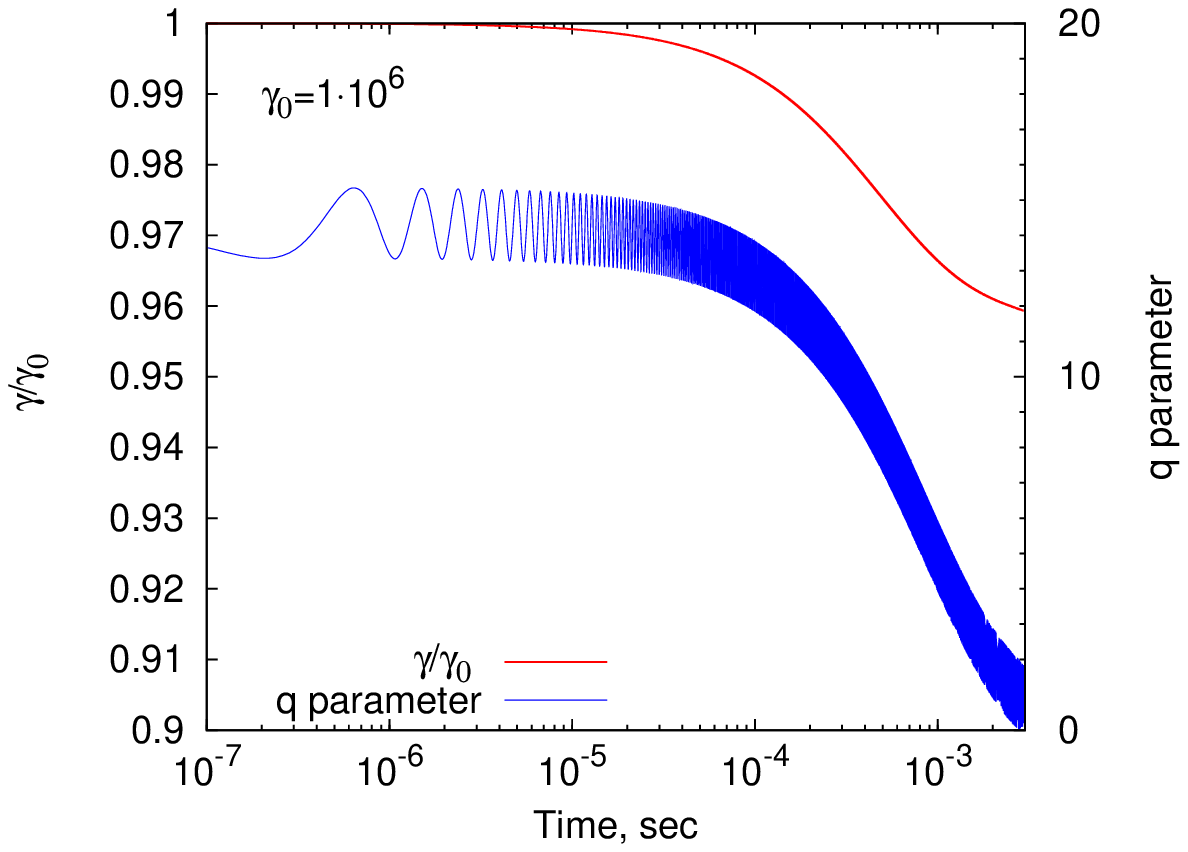} &
\includegraphics[width=0.25\textwidth]{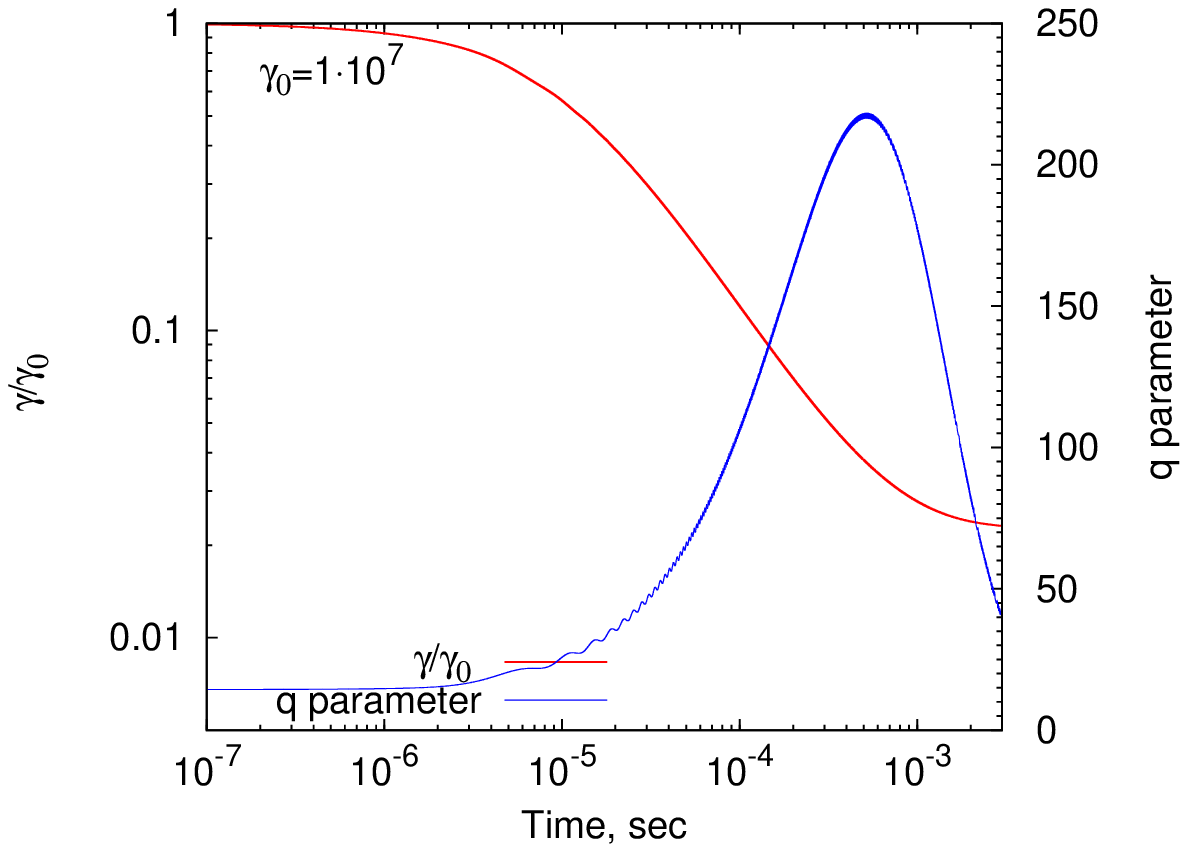} \\
\includegraphics[width=0.25\textwidth]{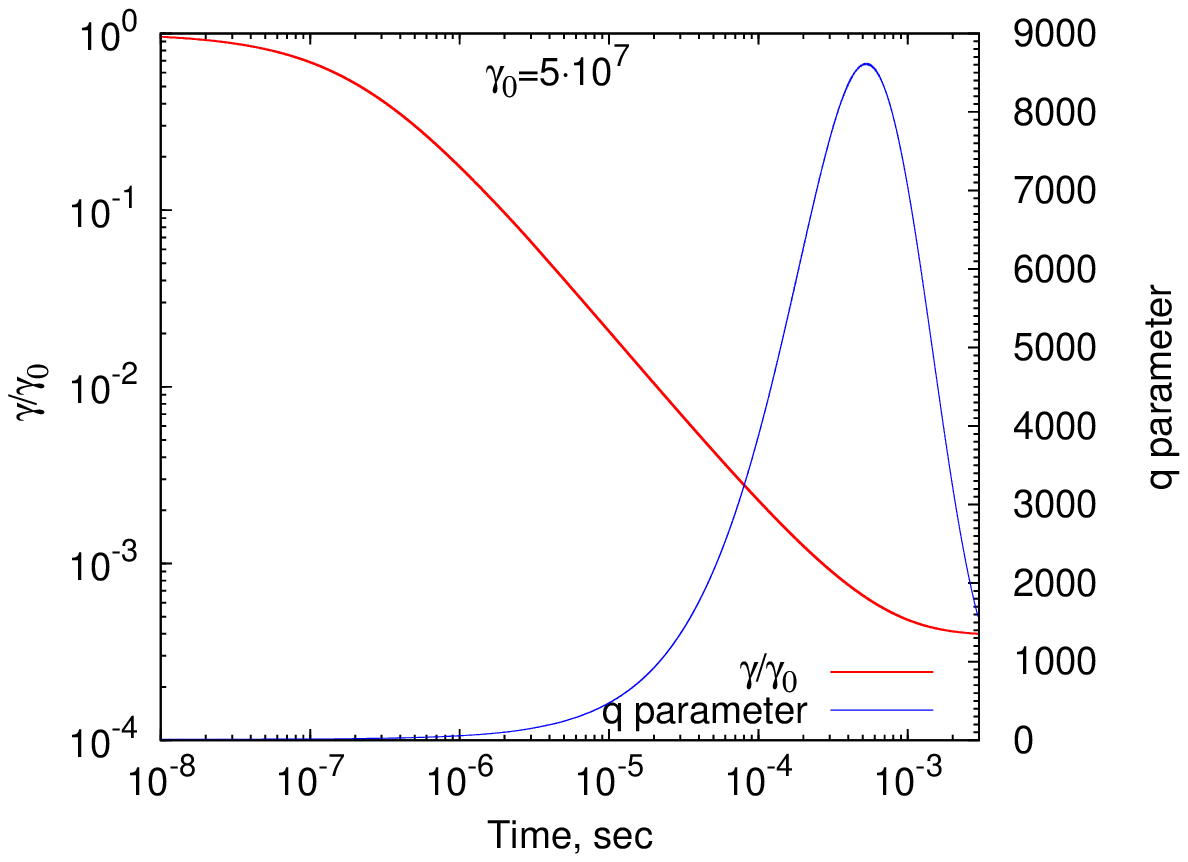} &
\includegraphics[width=0.25\textwidth]{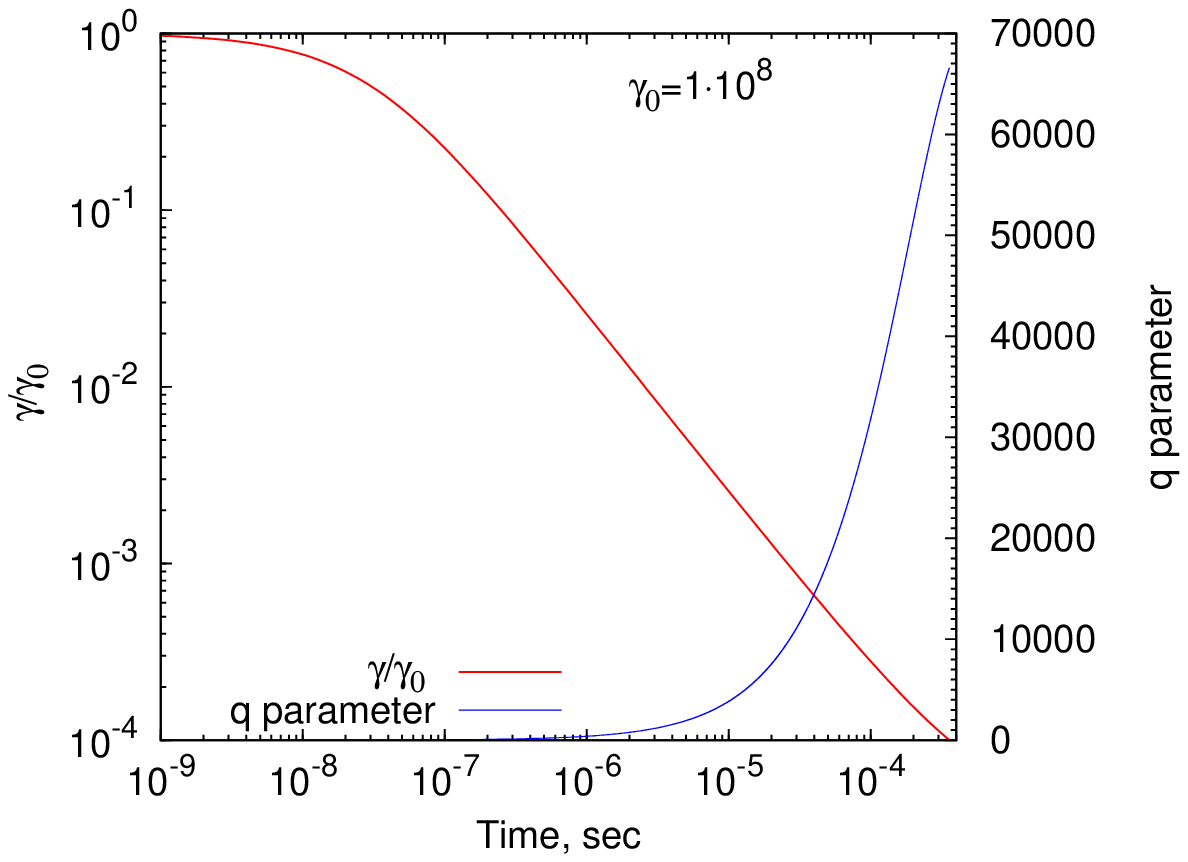}
\end{array}$
\end{center}
\caption{\label{fig:OGradComA}
Same as in Figure~\ref{fig:OGradComD}, but for the initial direction of electrons at the angle $10\beta_D$.
}
\end{figure}
The high-energy gamma radiation from pulsars is believed to
originate from the outer gap of the pulsar magnetosphere
\citep{Cheng1986,Takata2004,Hirotani2008}. Here we
present the results of our calculations of the radiation of
electrons (positrons) in the dipole magnetic field at the location
of the outer gap. The position of the outer gap is placed above
the last open field line. The inner boundary of the gap is at the
null surface where the Goldreich–Julian charge density is zero,
and the outer boundary is taken on the surface of the light
cylinder. We adopt the parameters of the Crab pulsar: the
radius of the star $R_*=10^6$ cm, the rotation period $P=33.5$ ms,
and the magnetic field at the pole $B_*=10^{12}$ G. We
consider a non-aligned pulsar with an angle of $45^{\circ}$ between the
rotation axis and the magnetic dipole axis. The initial particle
position is set at the distance $r_{\rm init}=0.5R_{lc}$ from the null surface
along the last open field line, where $R_{lc}$ is the radius of the light
cylinder. In numerical calculations, the particle is followed up
to the intersection with the light cylinder. For parameters
typical for the Crab pulsar, these conditions correspond to
$\theta_{0}=51.5^{\circ}$, $R_{0}=1.2\cdot 10^8$ cm, $B_{0}=5.6\cdot 10^{5}$ G, and
$R_{lc}=1.6\cdot 10^{8}$ cm, where $\theta_{0}$ is the polar angle relative to
the magnetic dipole axis.

The energy spectra and the radiation regimes were studied
for different initial directions and Lorentz factors of electrons.
The resulting spectra are shown in Figure~\ref{fig:OGrad}. The curves in
Figures~\ref{fig:OGradComD}--\ref{fig:OGradComA} demonstrate the time-evolution of the $q$-parameter
(blue lines, right scale) and the Lorentz factor normalized to its
initial value. These complementary plots allow us to simultaneously
watch the energy loss rate and the regime of the
radiation characterized by the value of the $q$-parameter. The
saltatory behavior of some curves for the $q$-parameter is
produced by small number of data points in the region when
the energy loss rate is low.

It follows from the above discussion that the particle with a
given energy radiates less intensively and produces less
energetic photons when it moves along drift trajectory (i.e.,
in the curvature radiation regime). Indeed, locally such motion
corresponds to the case when $\eta\ll 1$ or $\beta_{\perp}\ll \beta_D$ in
Equations~(\ref{cur38}), (\ref{cur42}), and (\ref{cur44}). It means that $q$ oscillates
around $1$ with a small amplitude of $\eta$. Then the peak of the
spectrum given by Equation~(\ref{cur44}) does not shift from the
position determined in curvature radiation. To check this
conclusion, we set the initial direction of the particle along the
drift trajectory, thus fixing the initial velocity at the angle $\beta_D$
from the magnetic field line toward the binormal vector. The
corresponding cumulative (integrated along the trajectory)
energy spectra of radiation are shown in Figure~\ref{fig:OGrad}. The
complementary plots are shown in Figure~\ref{fig:OGradComD}. We can see that,
indeed, for the initial Lorentz factors $\gamma_0=10^6$ and $\gamma_0=10^7$
the curvature radiation is less energetic compared to other cases
and the rates of energy losses are minimum as well (compare
with corresponding curves in Figures~\ref{fig:OGradComB} and \ref{fig:OGradComA}).

However, the situation is different for the larger initial
Lorentz factors $\gamma_0=5\cdot 10^7$ and $\gamma_0=10^8$ shown on the right
panel of Figure~\ref{fig:OGrad}. It is seen that for the initial direction along
the drift trajectory, the radiation spectra extend to higher
energies than in the case of the initial direction along the
magnetic field. This can be explained by the very intense
energy losses occurring before the particle has made the first
gyration (the first oscillation of $q$-parameter). During this time
interval, $q\approx 1$ for the initial direction along the drift trajectory
and $q\approx 0$ for the initial direction along the magnetic field.
Equations~(\ref{cur43}) and (\ref{cur44}) show that larger values of $q$ give a
more energetic radiation. Correspondingly, more energetic
radiation is produced in the case of the initial direction along
the drift trajectory. However, after many gyrations the energy
losses in the case of the initial direction along the drift
trajectory become, as expected, less intense compared to the
case of the initial direction along the magnetic field line.

In Figure~\ref{fig:OGrad} we compare the exact calculations of radiation in
the curvature regime with the results obtained under the
assumption that the electron moves strictly along the magnetic
field (as is routinely assumed in many papers on curvature
radiation). As expected, for the relatively modest initial Lorentz
factors $\gamma_0=10^6$ and $\gamma_0=10^7$, on the left panel of Figure~\ref{fig:OGrad},
the curves for these cases merge. For larger initial Lorentz
factors $\gamma_0=5\cdot 10^7$ and $\gamma_0=10^8$ the differences are not
negligible, but are still small.

We call the attention of the reader to the regimes of radiation
demonstrated by the curves in Figure~\ref{fig:OGradComD}. For $\gamma_0=10^6$ and
$\gamma_0=10^7$, the $q$-parameter equals unity, indicating that the
radiation proceeds in the curvature regime. However, this
equality is not exact, as is demonstrated for $\gamma_0=10^7$
where $q$-parameter slightly oscillates around unity. These small
oscillations correspond to the fine gyration around the drift
trajectory. As discussed above (see Equation (\ref{eq:applim})) the
presence of such fine perpendicular motions does not influence
on the applicability limits.

For $\gamma_0=5\cdot 10^7$ and $\gamma_0=10^8$, the $q$-parameter has a more
complex behavior. The increase at the beginning is defined
mostly by the fast energy losses. The decrease is determined by
the combination of several factors, such as the reduction of the
magnetic field strength and the change of its curvature. The
increase of $q$-parameter indicates that the radiation occurs in the
synchro-curvature or the synchrotron regimes when the
radiation due to curvature of the magnetic field line is less
important (for $\gamma_0=5\cdot 10^7$) or simply negligible (for
$\gamma_0=10^8$). According to Equation~(\ref{cur44}), the energy of the
radiation maximum scales as $\sim\gamma^3 q$. The $q$-parameter reaches
the maximum when a considerable amount of energy has been
lost. Therefore, in spite of large $q$, the peak of radiation shifts
toward low energies and does not affect the cumulative
spectrum. The interesting feature can be seen at first moments
when the energy oscillates with the $q$-parameter and the
minimums of $q$-parameter correspond to flatter parts of the
Lorentz factor evolution curve.

The radiation spectra of electrons launched along the
magnetic field line are slightly more energetic, except for
$\gamma_0=5\cdot 10^7$ and $\gamma_0=10^8$ as discussed above. Initially, the
radiation for $\gamma_0=10^6$ and $\gamma_0=10^7$ is in the synchro-curvature
regime (see Figure~\ref{fig:OGradComB}), although for $\gamma_0=10^6$ the most of the
energy is lost in the curvature regime, which occurs fast.
Therefore the spectrum in this case almost coincides with the
spectrum for the initial direction along the drift trajectory. For
$\gamma_0=10^7$ most of the energy is lost in the synchro-curvature
regime, thus the spectrum is shifted to higher energies by a
factor of $q\approx 2$.

For illustration of the effect related to initial pitch angles
larger than $\beta_D$, we show the case of the initial direction
deflected at the angle $10\beta_D$ from the magnetic field line toward
the direction opposite the normal vector. The corresponding
spectra indicated in Figure~\ref{fig:OGrad} by dashed-dotted lines are shifted
toward higher energies. Although the $q$-parameter (Figure~\ref{fig:OGradComA})
reaches large values, most of the energy is lost at initial stages
at $q\approx 10$. Thus the spectra are shifted by $q\approx 10$ compared to
the curvature radiation spectra.

We should note that the energy spectra of radiation produced
in all regimes at highest energies contain an exponential cut-off
similar to the spectrum of the small-angle synchrotron radiation.
However, since the radiation spectrum is very sensitive to the
pitch-angle, a population of electrons with similar energies but
different angles can result in a superposition spectrum with a less
abrupt cut-off. The condition for the realization of such a
spectrum is that the distribution of electrons over the angles
around the drift trajectory should be wider than~$\beta_D$.

\subsection{Polar Cap}
In the polar cap model the electron radiates in the region
located close to the surface of the neutron star, where the
magnetic field is much stronger than in the outer gap model,
approaching to $B\approx 10^{12}$ G. This results in much faster
damping of the perpendicular component of motion. The very
small drift velocity $\beta_D$ implies that the drift trajectory and the
magnetic field line almost coincide, thus even a small
deflection from the magnetic field line produces radiation that
is quite different from the curvature radiation. However, the
transition to the curvature radiation regime occurs very fast. In
the curvature regime electrons radiate more energetic photons
than in the outer gap model because of the stronger curvature of
the field lines. The curvature of the dipole field lines behaves
like $\sim\sin\theta/r$, and with the decrease of the radius by two orders
of magnitude and the decrease of the polar angle by an order of
magnitude ($\theta_0\sim 1^{\circ}$), the curvature is larger by an order of
magnitude compared to the case of outer gap model.
Correspondingly, the maximum energy of the curvature
radiation in the polar cap is an order of magnitude higher than
in the outer gap.

\begin{figure}
 \begin{center}
  \includegraphics[width=0.5\textwidth]{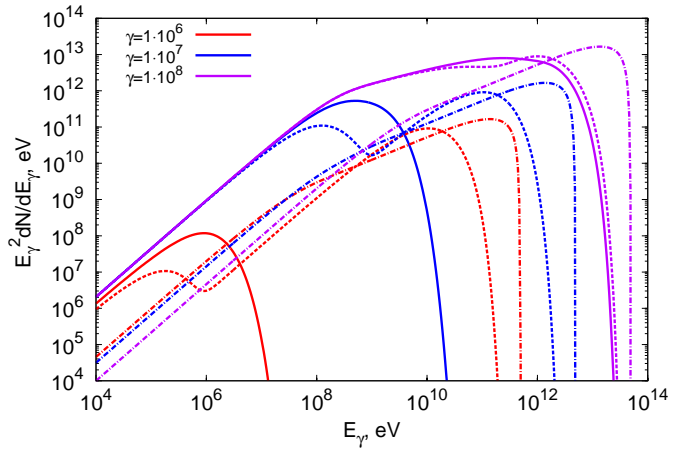}
  \caption{\label{fig:PCrad}
  Cumulative (integrated along trajectory) radiation spectra of
  electrons calculated for the polar cap model in the pulsar magnetosphere.
  The curves are obtained for different initial Lorentz factors of electrons
  $\gamma=10^6, 10^7, 10^8$, and for different initial directions relative to the magnetic
  field lines: along the magnetic field line (solid lines), and for two pitch angles
  $1/\gamma_0$ (dashed lines) and $100/\gamma_0$ (dot-dashed lines).
  }
 \end{center}
\end{figure}

\begin{figure}
\begin{center}$
\begin{array}{cc}
\includegraphics[width=0.25\textwidth]{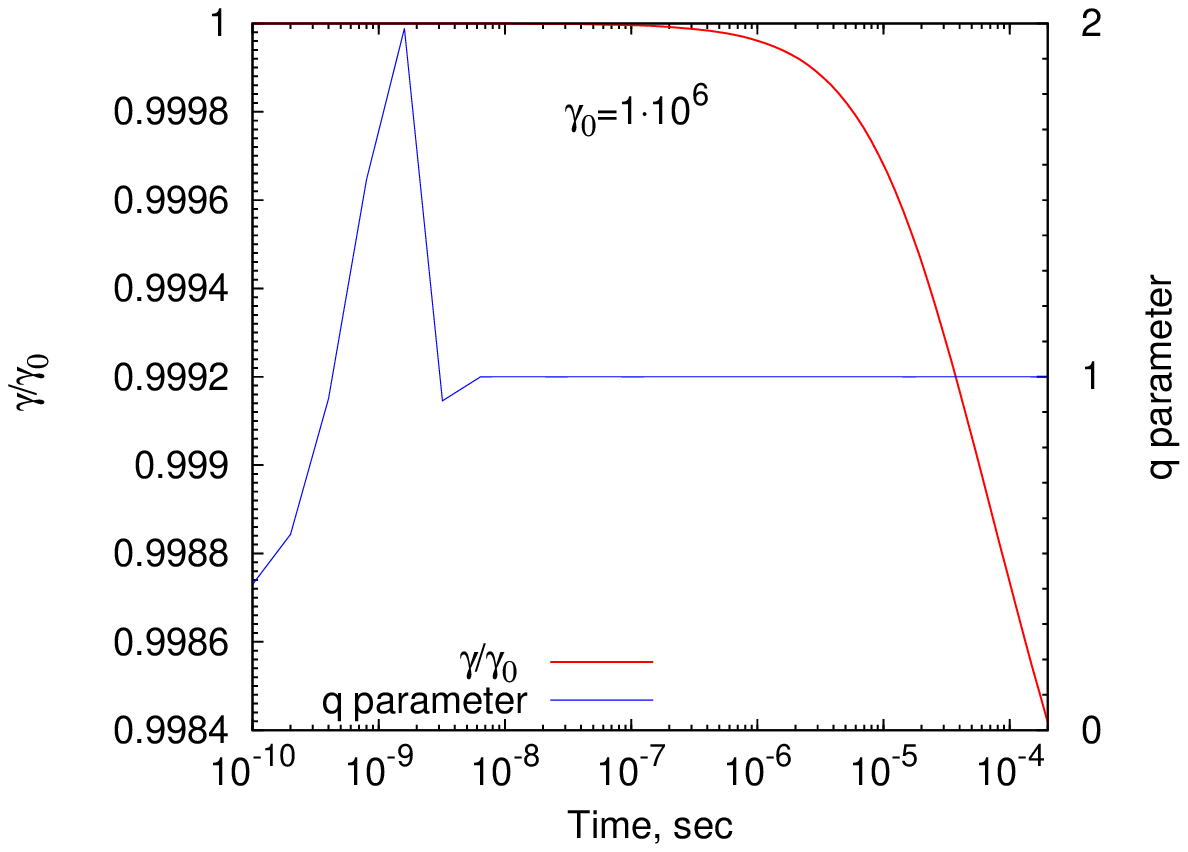} &
\includegraphics[width=0.25\textwidth]{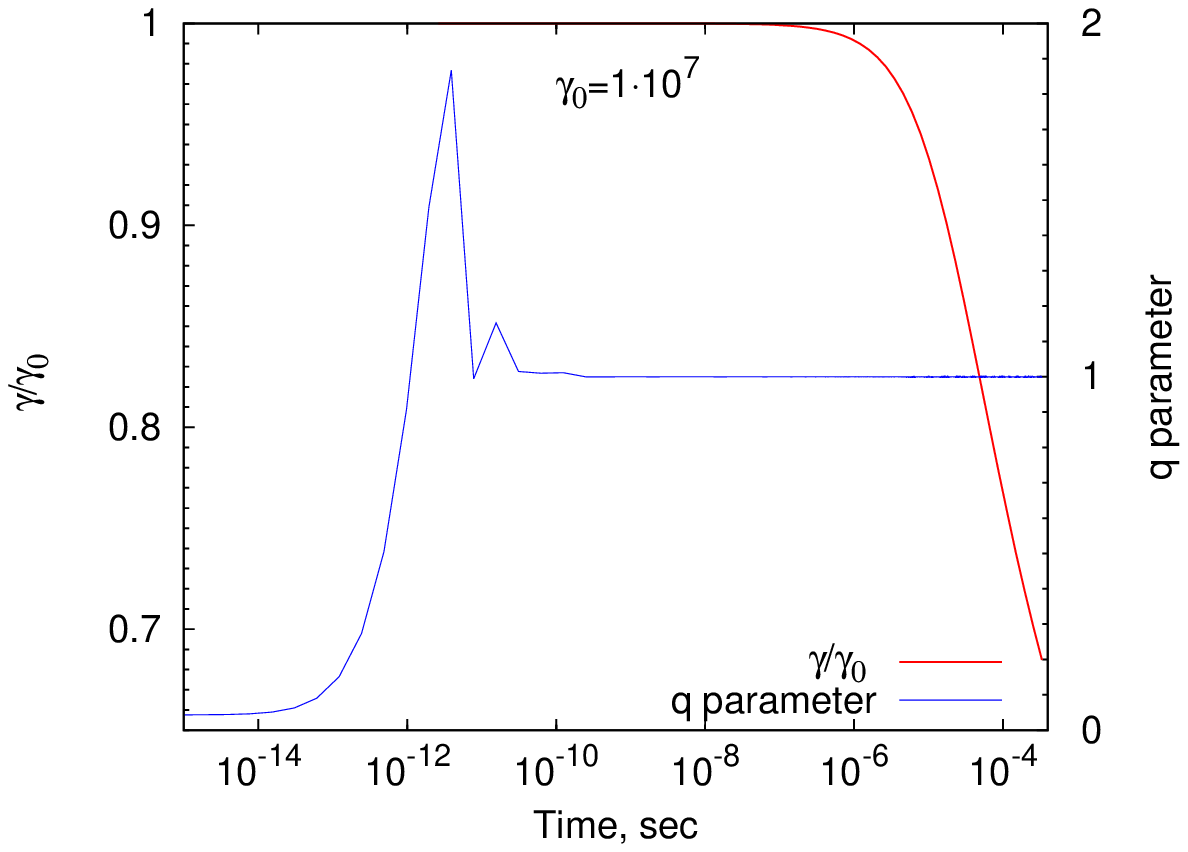}
\end{array}$
\includegraphics[width=0.25\textwidth]{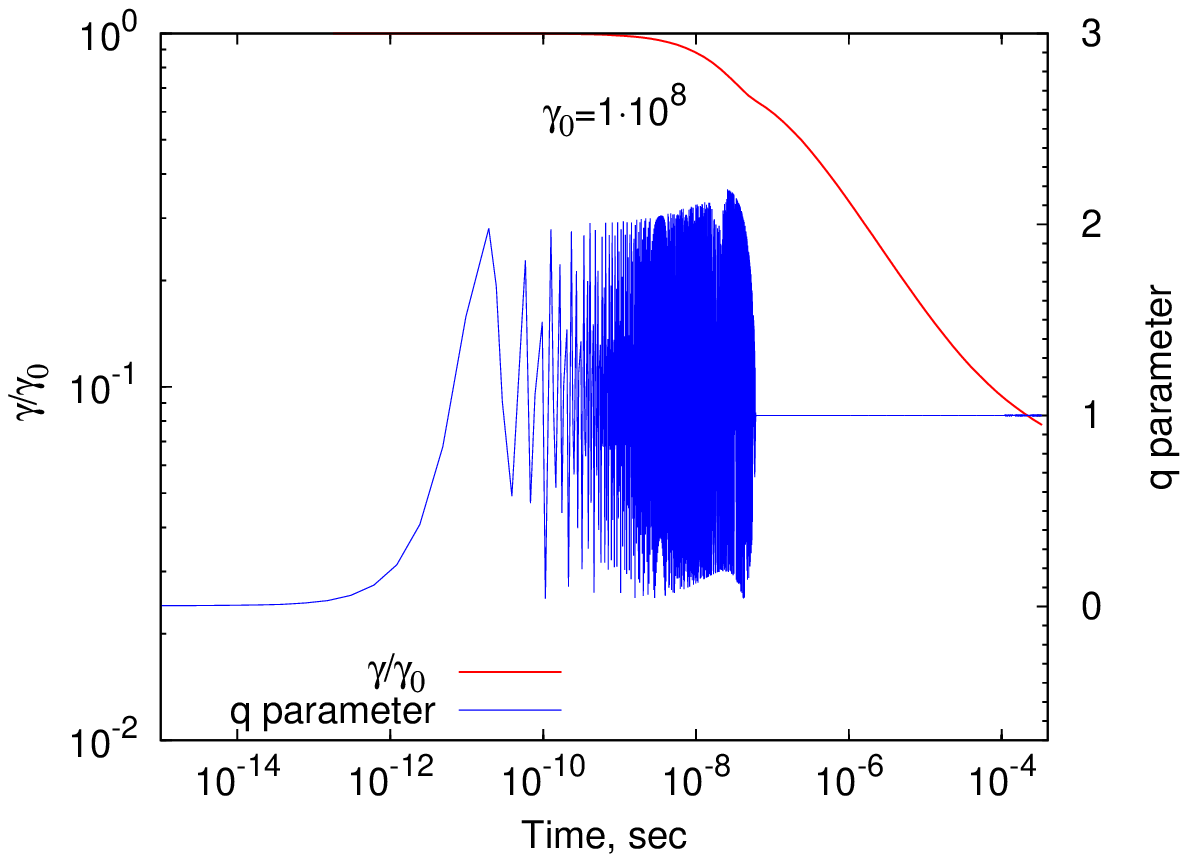}
\end{center}
\caption{\label{fig:PCalong}
 Time evolution of the $q$-parameter and the electron Lorentz factor in the outer gap model 
 (complementary to Figure~\ref{fig:PCrad}). The three panels correspond to the
 initial Lorentz factor of electrons $\gamma=10^6, 10^7,  10^8$ and their initial direction 
 along the magnetic field line.
 }
\end{figure}

\begin{figure}
\begin{center}$
\begin{array}{cc}
\includegraphics[width=0.25\textwidth]{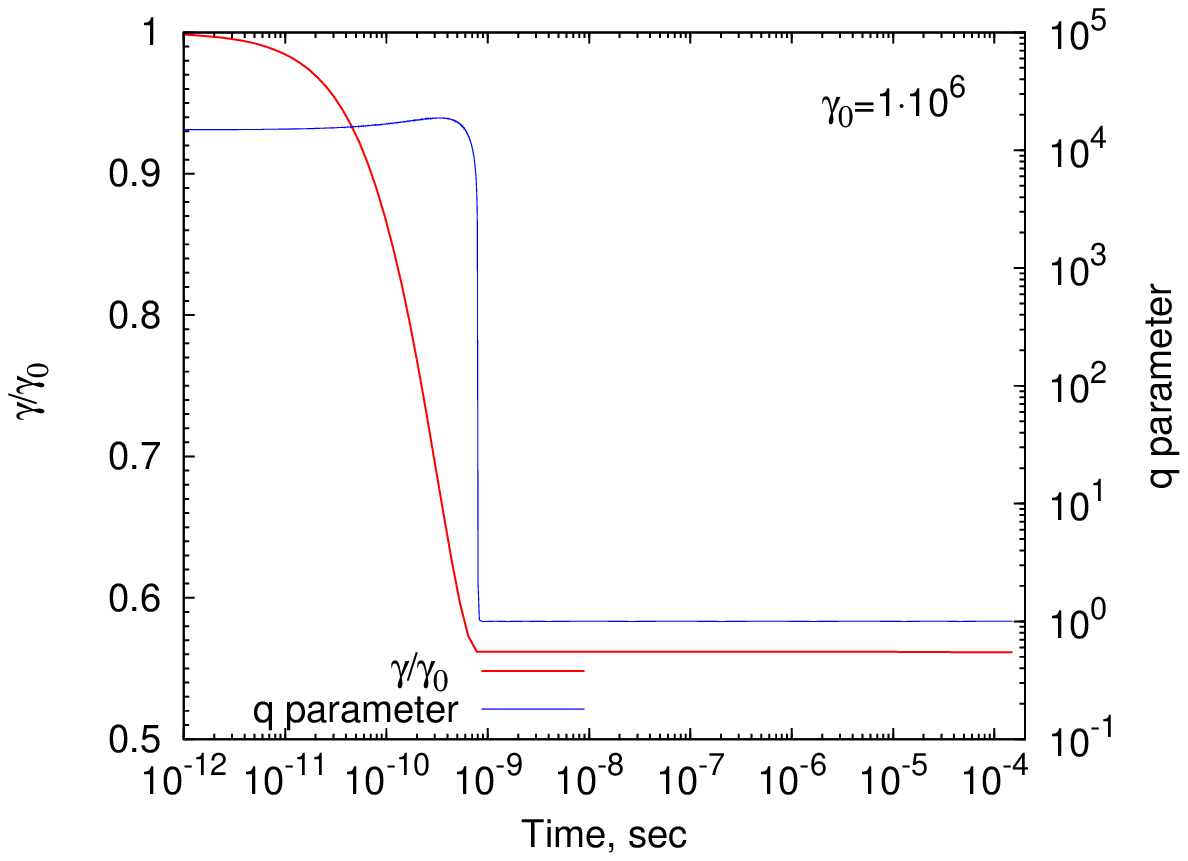} &
\includegraphics[width=0.25\textwidth]{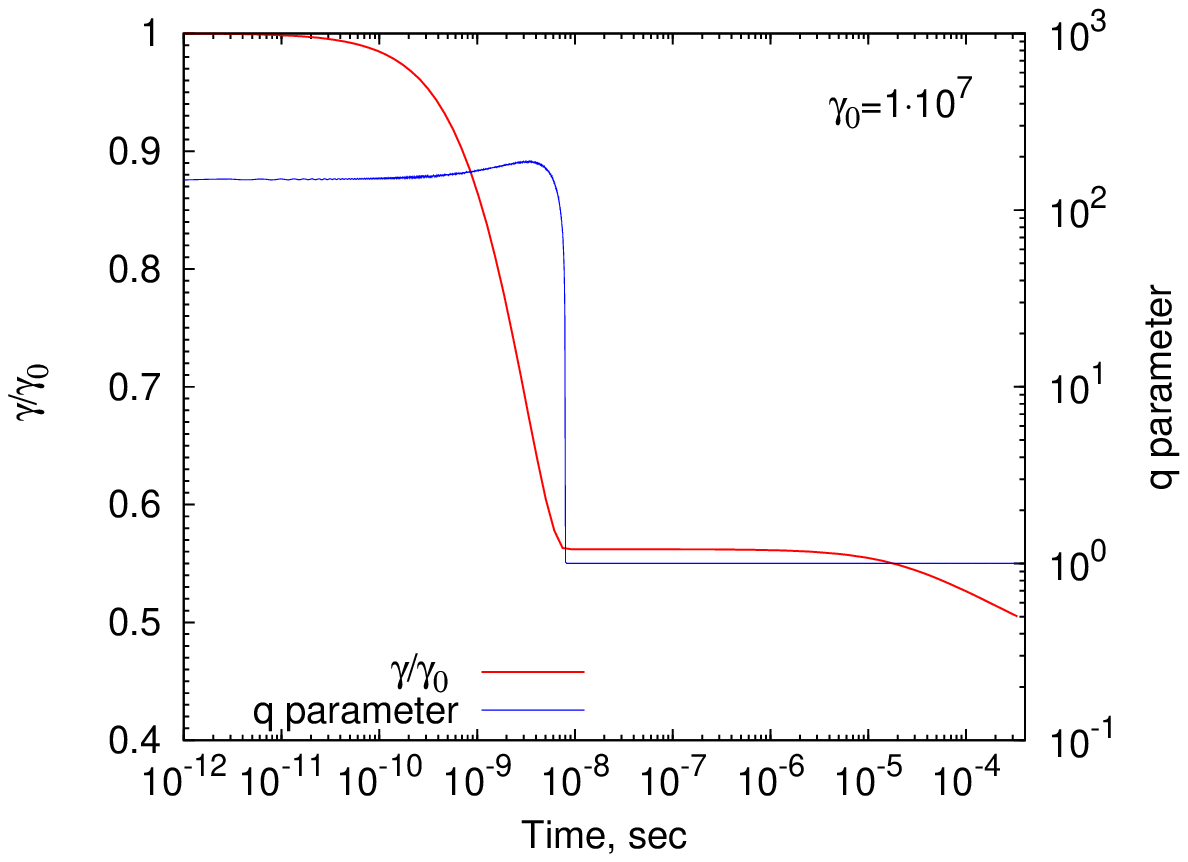}
\end{array}$
\includegraphics[width=0.25\textwidth]{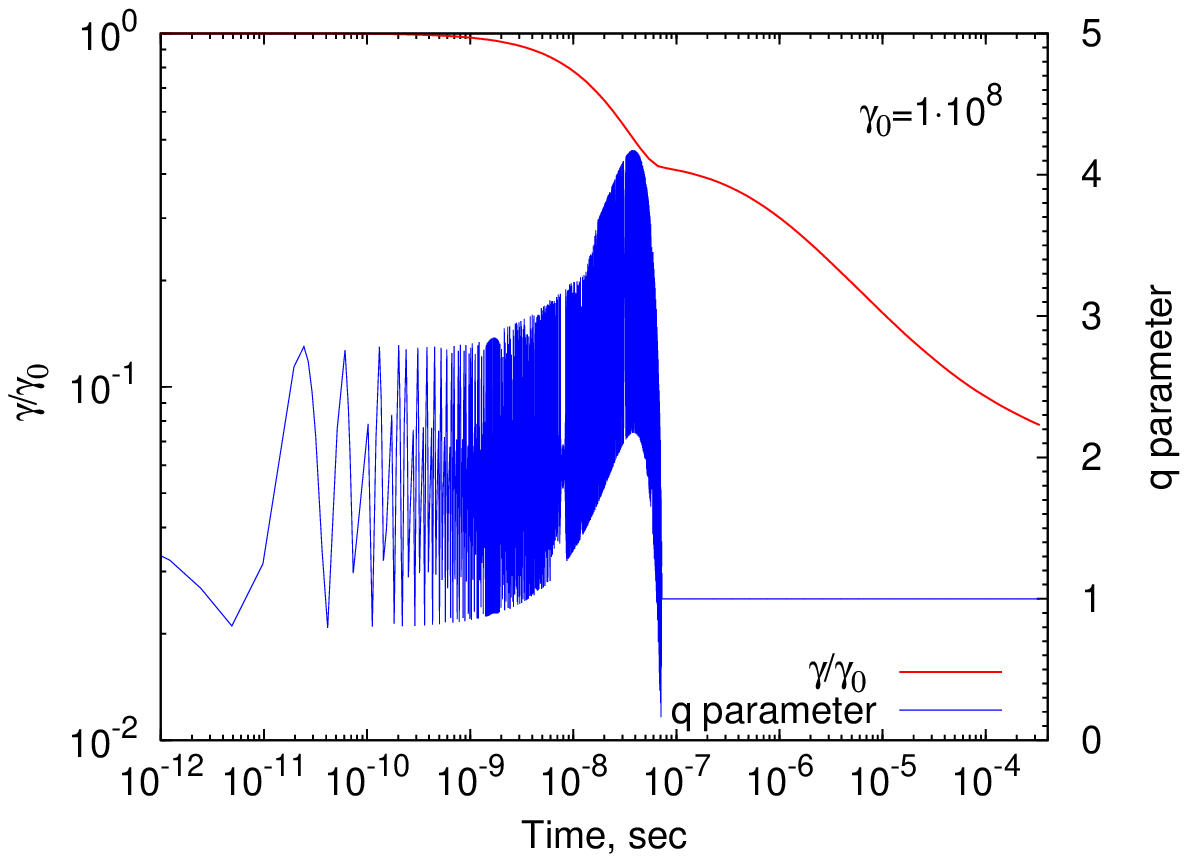}
\end{center}
\caption{\label{fig:PC1g}
Same as in Figure~\ref{fig:PCalong}, but for the initial direction of electrons at the pitch angle 
$1/\gamma_0$.
}
\end{figure}

\begin{figure}
\begin{center}$
\begin{array}{cc}
\includegraphics[width=0.25\textwidth]{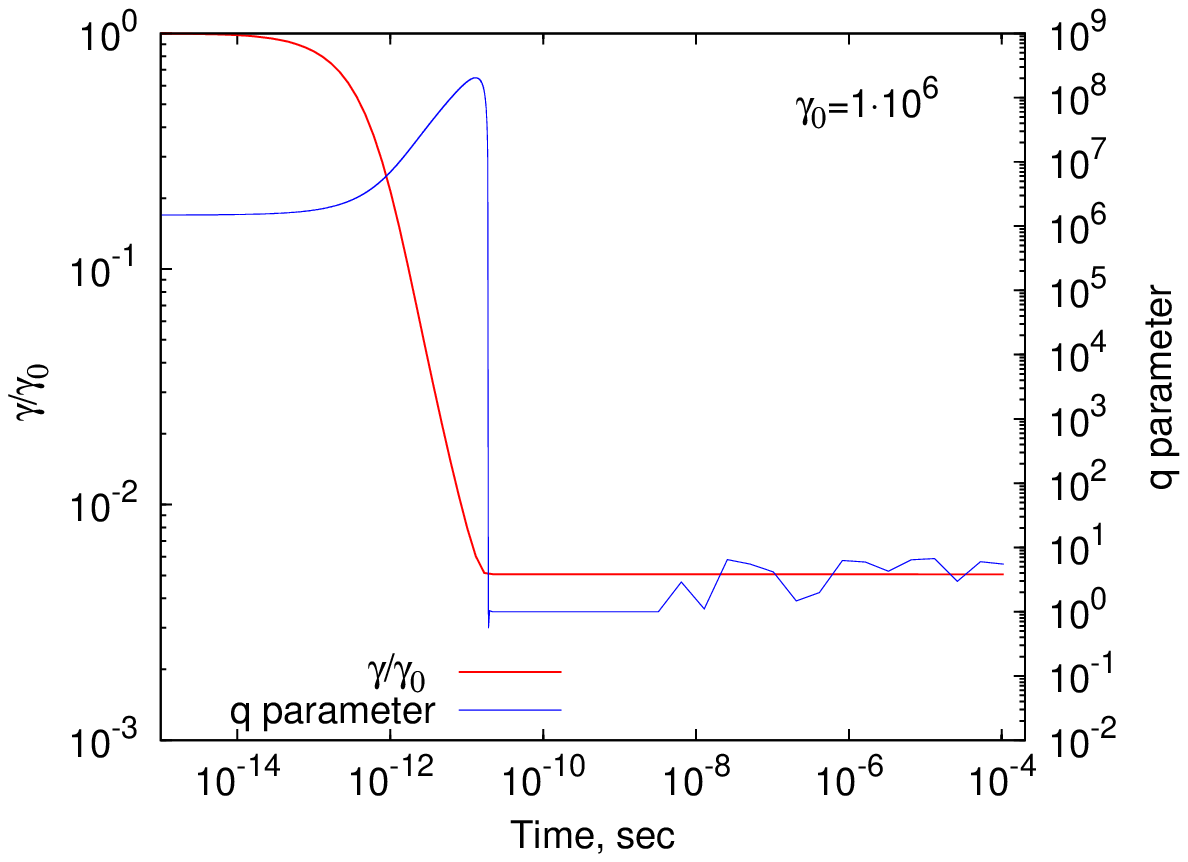} &
\includegraphics[width=0.25\textwidth]{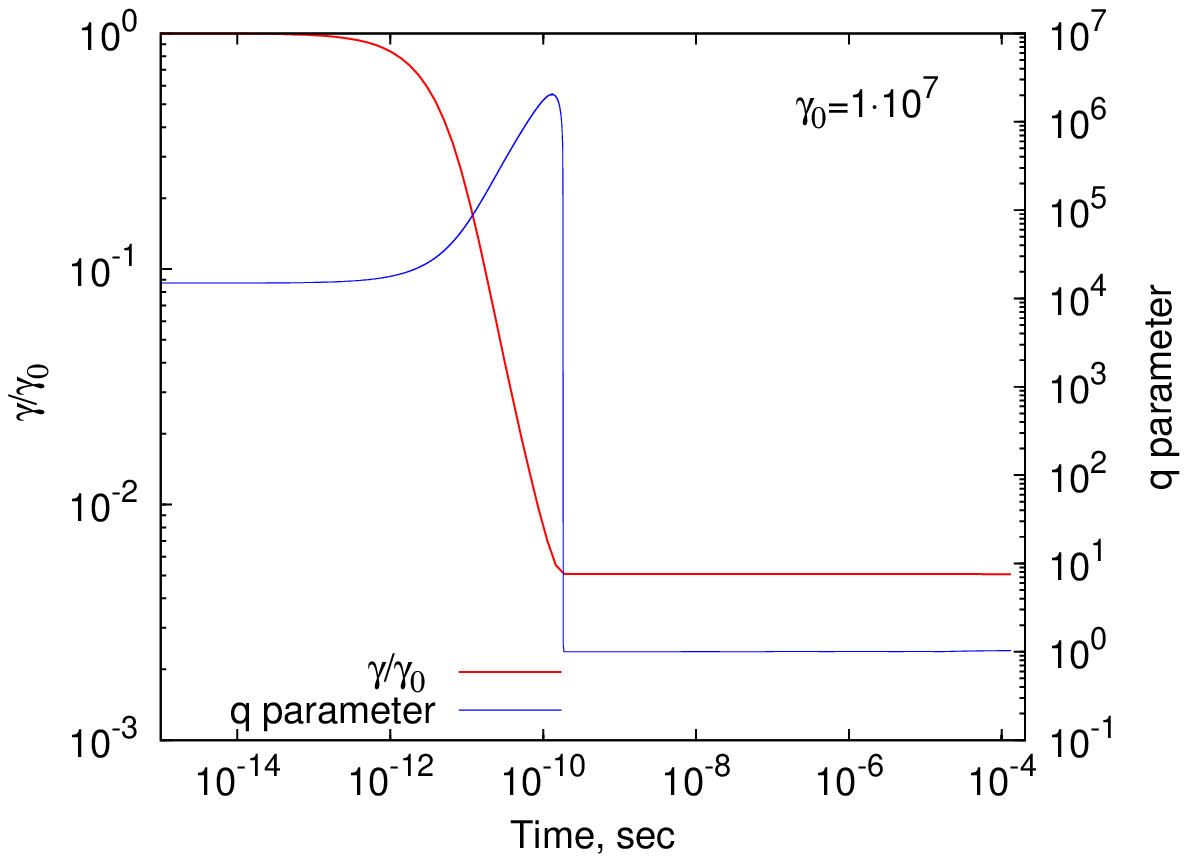} 
\end{array}$
\includegraphics[width=0.25\textwidth]{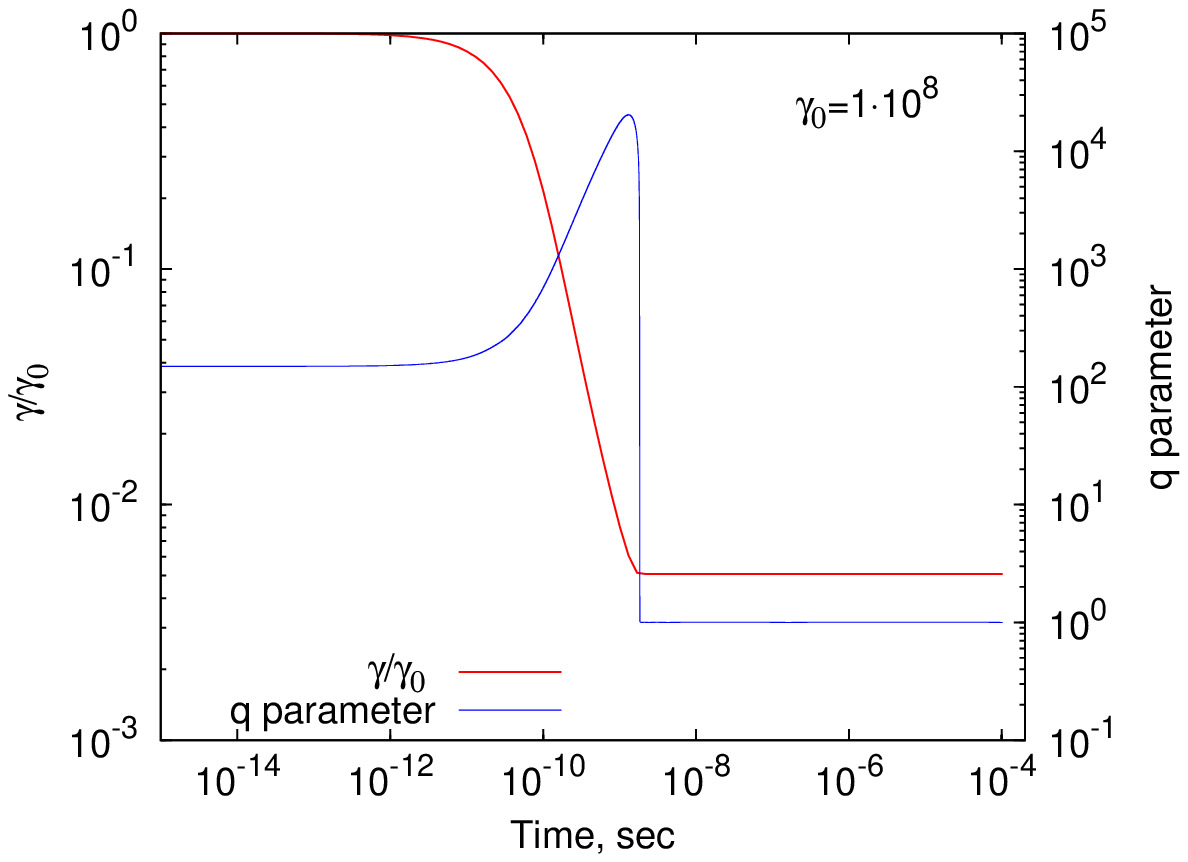}
\end{center}
\caption{\label{fig:PC100g}
Same as in Figure~\ref{fig:PCalong}, but for the initial direction of 
electrons at the pitch angle $100/\gamma_0$.
}
\end{figure}

The energy spectra of radiation calculated for the polar cap
model are shown in Figure~\ref{fig:PCrad}. The complementary plots for the
evolution of the $q$-parameter and the electron Lorentz factor are
presented in Figures~\ref{fig:PCalong}-\ref{fig:PC100g}. The initial position of the particle is
$R_0=10^6$ cm, and $\theta_0=3^{\circ}$. The spectra indicated by solid lines
correspond to the case when the initial direction of the particle
is along the magnetic field line. In this case the radiation is
in the curvature regime. Initially, the radiation shortly proceeds in
the synchro-curvature regime, but abruptly turns to the regime
with fast oscillations around $q \approx 1$ caused by fine gyrations
(Figure~\ref{fig:PCalong}).

In the case of an initial pitch angle $\alpha_0=1/\gamma_0$ the abrupt
change of regimes leads to an interesting feature in the
cumulative spectra (dashed lines). Because of small changes of
the energy and fast changes of the $q$-parameter, an energy
spectrum is formed consisting of two peaks. The peak at higher
energies is produced in the synchrotron regime ($q\gg 1$, see
Figure~\ref{fig:PC1g}), while the lower energy peak is due to radiation in
the curvature regime. The double-peak structure disappears for
large initial pitch angles. For example, for the pitch angle
$100/\gamma_0$ the transition to the curvature regime is very fast and the
electron enters into this regime with dramatically reduced
Lorentz factor. Thus the peak of the curvature radiation is not
only shifted to smaller energies, but also too weak to be seen in
the cumulative spectrum\footnote{Note that the pitch angle $100/\gamma_0$, which we treat as large, is still
extremely small, $\sim 2 (\gamma_0/10^7)^{-1}$ arcsec.}.

In very strong magnetic fields, namely when the parameter
$\chi=B\gamma \sin\alpha/B_{cr} \geq 1$, the radiation is produced in the
quantum regime. Let's assume that the initial pitch angle is
inversely proportional to the initial Lorentz factor, $\alpha_0=a/\gamma_0$.
This makes the condition of radiation in the quantum regime
independent of $\gamma_0$:
  
\begin{equation}
\chi=\frac{B}{B_{cr}}\gamma \sin{\alpha_0}=\frac{1\cdot 10^{12} G}{2.94\cdot 10^{13} G} a \approx 3.4\cdot 10^{-2} a.
\end{equation}
Thus, at the initial pitch angle with $a>30$, the electrons
radiate in the quantum regime. Dashed-dotted lines in Figure~\ref{fig:PCrad}
present radiation spectra for the initial pitch-angles
$\alpha_0=100/\gamma_0$. Note that in the quantum regime almost the
entire energy of the parent electron is transferred to the radiated
photon. Therefore we should expect an abrupt cutoff in the
radiation spectra. This effect is clearly seen in Figure~\ref{fig:PCrad} (dot-dashed 
curves corresponding to the initial pitch angle
$\alpha_0=100/\gamma_0$). The gamma-rays produced in the quantum
regime are sufficiently energetic to be absorbed in the magnetic
field through the $e^+e^-$ pair production. This leads to the
development of an electromagnetic cascade in the magnetic
field. The spectrum of cascade gamma-rays that escape the
pulsar magnetosphere will be quite different from the spectra
shown in Figure~\ref{fig:PCrad}.

\section{Motion and radiation in electric and magnetic fields}\label{sec:elec}
To study a more realistic picture of the radiation, one should
consistently determine the initial pitch angle. This can be done
if the acceleration of the particle is considered. First we
consider the motion of a charged particle in the crossed
homogeneous electric and magnetic fields. If the fields are not
perpendicular to each other, one can always find a reference
frame in which they are parallel. In this reference frame the
particle completely loses the momentum perpendicular to the
common direction of the electric and magnetic fields. At the
same time, the electric field infinitely accelerates the particle.
Eventually the particle accelerates along the common direction
of the electric and magnetic fields. In an arbitrary reference
frame the particle is accelerated along a rectilinear trajectory.
However, the direction of the motion is

\begin{equation}\label{eq:DirAcc}
\b \beta=\frac{1}{Q} \mathbfcal{E}\times\b B \pm\left(\frac{\sqrt{Q-\b B^2}}{Q}\mathbfcal{E}+\frac{\sqrt{Q-\mathbfcal{E}^2}}{Q}\b B\right),
\end{equation}
where 
\begin{equation}
Q=\frac{\mathbfcal{E}^2+\b B^2}{2}+\sqrt{\left(\frac{\mathbfcal{E}^2-\b B^2}{2}\right)^2+(\mathbfcal{E}\b B)^2},
\end{equation}
and $+$ and $-$ signs correspond to positively and negatively
charged particles, respectively. Then the pitch angle can be
expressed as
\begin{equation}\label{eq:pitchang}
\sin \alpha=\frac{\sigma \sin
\theta}{\sqrt{\frac{1}{2}\left(\sqrt{(1+\sigma^2)^2-4\sigma^2\sin^2\theta}+(1+\sigma^2)\right)}},
\end{equation}
where $\sigma=\mathcal{E}/B$ is the ratio of the electric and magnetic fields,
and $\theta$ is the angle between them. Assuming that the electric
field is typically smaller than the magnetic field, we obtain
\begin{equation}\label{eq:electdr}
\alpha\approx\beta_{E}=\sigma \sin\theta\ll 1
\end{equation}
which implies that the pitch angle equals the electric drift
velocity (in units of the speed of light). Thus, in the crossed
homogeneous electric and magnetic fields, the asymptotic
motion of the particle after damping of gyration is a constant
acceleration along the rectilinear trajectory. The direction of the
trajectory constitutes the angle $\beta_E$ with the direction of the
magnetic field. Note that due to this acceleration the particle
radiates. However, the energy losses do not depend on the
Lorentz factor and are negligibly small compared to the energy
losses during the particle motion along the curved trajectory.

In the curved electric and magnetic fields the acceleration of
the particle is restricted by the energy losses due to the curved
trajectory. Once the characteristic times of the acceleration and
radiation become equal, the particle moves along a drift
trajectory determined by the electric and magnetic fields.
During radiation the particle changes its direction of motion. At
the same time it is accelerated by the electric field in the
direction determined by Equation~(\ref{eq:DirAcc}). As in the case of
crossed homogeneous electric and magnetic fields, the
component of the momentum perpendicular to the drift
trajectory, which is responsible for the gyration of the particle,
is damped due to radiation. Thus, in the equilibrium between
acceleration and radiation, the particle moves along drift
trajectory practically without gyration with the pitch angle $\beta_{E}$.
If the electric or magnetic field changes slowly along trajectory,
it gradually adjusts itself to the equilibrium state of the drift
trajectory. However, if fields vary fast, it can lead to a strong
radiation until the particle gets a new equilibrium drift
trajectory. Such conditions might appear when the particle
escapes the acceleration gap and the field-aligned (along
magnetic field lines) electric field is screened.

\begin{figure}
\begin{center}
\includegraphics[width=0.5\textwidth]{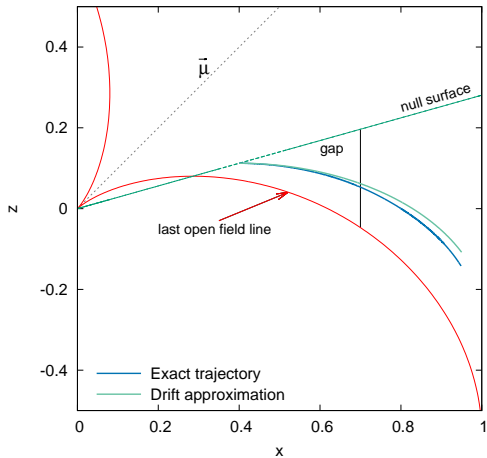}
\end{center}
\caption{\label{fig:traj} 
Projection on the XZ plane of the trajectories of the electrons
emitting the spectra in Figure~\ref{fig:accel}. The initial Lorentz factor is $\gamma_0=100$. The
particles begin to be accelerated from the null surface at $x=0.4R_{lc}$. The
distances of axes in units of the light cylinder. The border of the gap is at
$x=0.7R_{lc}$. The trajectories outside the gap are presented for the case of
exponential screening with the width $W=0.1R_{lc}$.
}
\end{figure}

Further we study the radiation of the particle inside and
outside the outer gap. We check how the character of the
radiation changes depending on the variation of the electric
field. Specifically, we consider the case of screening the fieldaligned
electric field when it decreases exponentially from the
border of the outer gap. The geometry of the model is presented
in Figure~\ref{fig:traj}. As before, the angle between the rotation and
dipole axes is $45^{\circ}$. The strength of the magnetic field at the pole
of the star is $B_*=10^{12}$ G. The period of rotation of the star is
$P=33$ msec. The configuration of the electric and magnetic
fields is given by the approximation of \cite{Deutsch1955} for the
fields in the vacuum in the vicinity of a rotating star (equations
(18) and (19) in \cite{Deutsch1955}). The magnetic field in this
approximation is just a dipole magnetic field. The outer gap is
located between the null surface and the last open field line. We
set the border of the gap at $x_{b}=0.7R_{lc}$ from the rotation axis
(see the black vertical line in Figure~\ref{fig:traj}), where $R_{lc}$ is the radius
of the light cylinder. In our calculations we assume that the
field-aligned electric field is screened starting from $x_{b}$
according to the exponential decay:

\begin{equation}
\mathcal{E}_{\parallel}=\mathcal{E}_{0\parallel}e^{-\left(\frac{x-x_{b}}{W}\right)^2},
\end{equation}
where $\mathcal{E}_{0\parallel}$ is the original component of the electric field parallel
to the magnetic field (as it would change without screening).
Note that the perpendicular component stays unmodified.
Changing the width of the screening $W$, we can explore
different models for the variation of the electric field. The three
cases that we investigate, $W=10^{-3}R_{lc}$, $10^{-2}R_{lc}$, and $10^{-1}R_{lc}$,
cover a large range of values of this parameter. The calculation
of the trajectory starts from the null surface at a distance
$x=0.4R_{lc}$ from the rotation axis, with initial Lorentz factor
$\gamma_0=100$ and velocity directed along magnetic field line.

Together with calculations using the exact equations of
motion given by Equation~(\ref{eq:system}), we performed the calculations
in the drift approximation presented by Equation~(\ref{eq:systemApp}). The
approach of the drift approximation has been used in the work
\cite{Kalapotharakos2012} to calculate the light curves for
the different models of the magnetosphere. Here it is important
for us to investigate how well this approach reproduces the
radiation spectra. Apart from the described configuration of the
electric field, we carried out the calculations for more simple
electric fields when their strength is proportional to the strength
of the magnetic field and their direction is always at a constant
angle to the direction of the magnetic field at a given point. The
results for a broad range of values for the angles and strength
ratios have shown a very good agreement between the exact
and approximate approaches. We do not show them because
the case of the configuration considered here is quite
representative. The explanation of the good agreement between
radiation spectra calculated in exact and approximate
approaches is that in both cases the curvatures of the
trajectories are very close. As discussed previously, in the
equilibrium between the acceleration and radiation the particle
basically moves along drift trajectory, which is assumed
a priori in the drift approximation.

The projection of the trajectories on the XZ plane calculated
in both approaches are presented in the Figure~\ref{fig:traj}. The
trajectories outside the gap are shown for the case of the slow
variation of the electric field $W=10^{-1}R_{lc}$ when the scale of
change is of the same order as the scale of the gap. Before we
discuss the influence of screening on the particle motion and
radiation, let us consider phenomena in the gap.

\begin{figure*}
\begin{center}
\mbox{\includegraphics[width=0.5\textwidth]{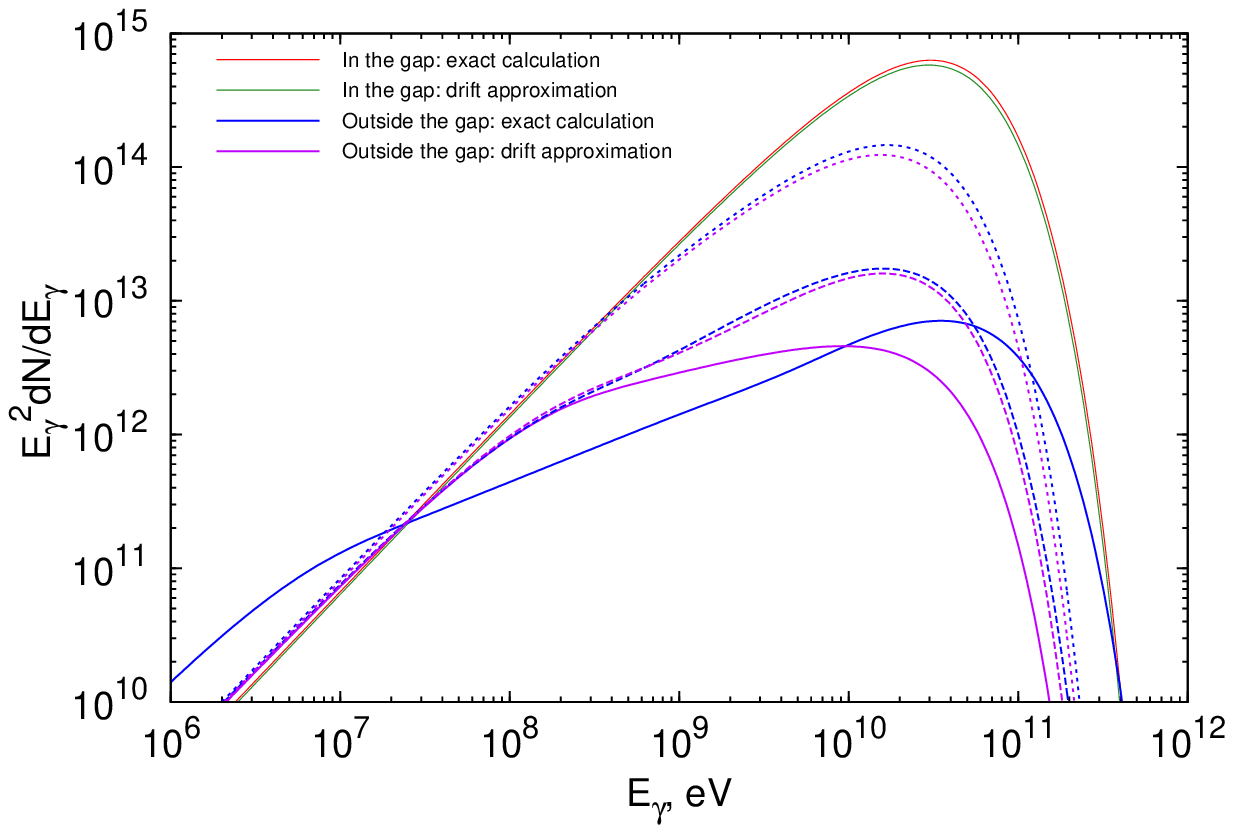}
\includegraphics[width=0.5\textwidth]{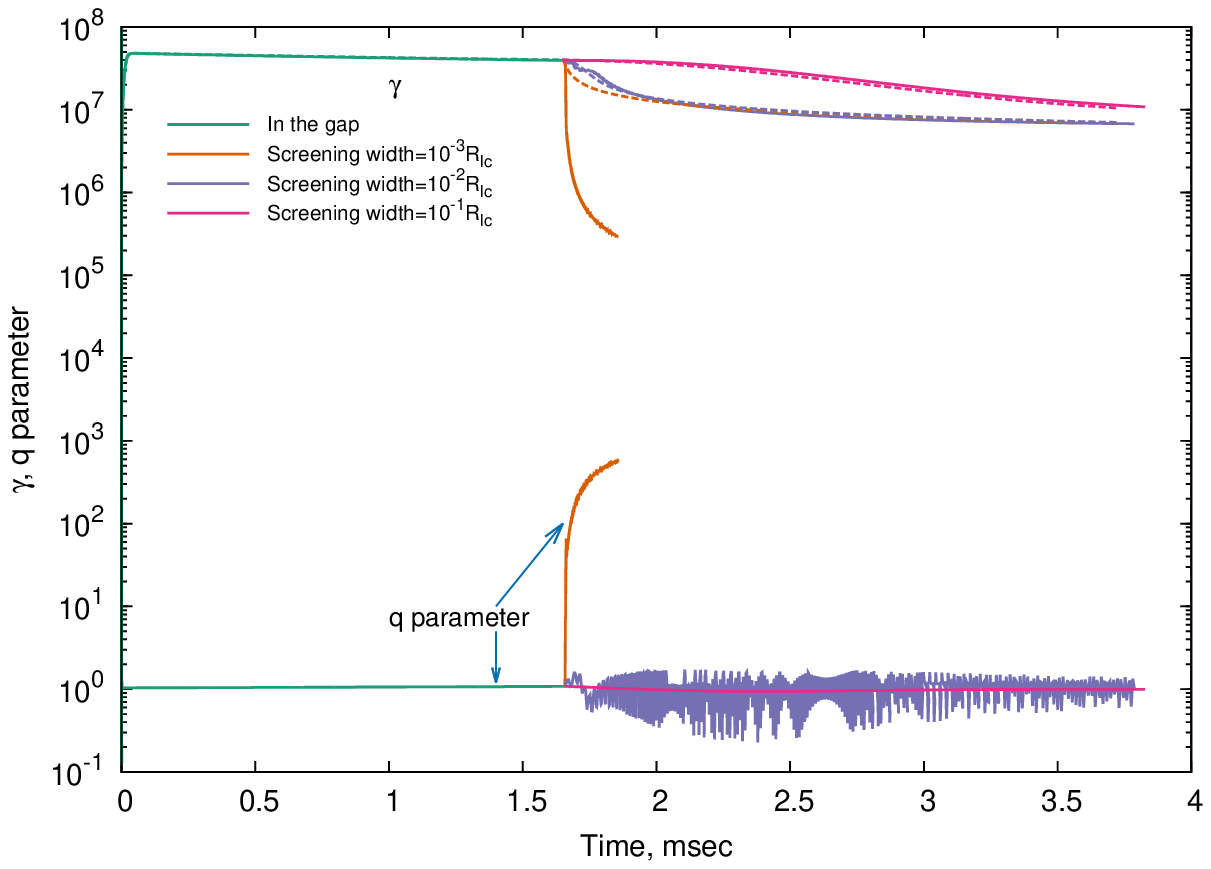}}
\caption{\label{fig:accel}
Left panel: cumulative radiation spectra of electrons accelerated in the gap (red and green lines), and radiation spectra of electrons outside the gap (blue
and purple lines) with screening width $W=10^{-1}R_{lc}$ (dotted lines), $W=10^{-2}R_{lc}$
(dashed lines), $W=10^{-3}R_{lc}$ (solid lines) obtained in the exact numerical
calculations (blue lines) and using the drift approximation (purple lines). Right panel: evolution of the Lorentz factor and $q$-parameter with time.
 }
\end{center}
\end{figure*}

Aside from the initial moments when the particle is
accelerated fast, it moves along equilibrium drift trajectory.
In the equilibrium regime when the acceleration by electric
field is balanced by energy losses of radiation, the position of
the maximum of the energy distribution $\epsilon_{*}$ is given by (see
Appendix~\ref{sec:maxen}):
\begin{equation}\label{eq:maxofen}
\epsilon_{*}=\left(\frac{3}{2}\right)^{\frac{7}{4}}\left(\frac{\sigma\cos\theta B}{e}\right)^{\frac{3}{4}}c\hbar\sqrt{r_0
\left(1+\sigma^2\sin^2\theta\right)},
\end{equation}
where $r_0$ is the curvature of the magnetic field line, and $e$ is the
elementary charge. The equation is valid for $\sigma<1$. Note that in
the derivation of Equation~(\ref{eq:maxofen}), the energy losses were taken in
the form given by Equation~(\ref{cur36}). For comparison, if the
cooling of electrons is dominated by synchrotron losses, we
obtain the classical limit for maximum energy of radiation:

\begin{equation}\label{eq:maxclass}
\epsilon_{syn}=\frac{9}{4}\frac{mc^2}{\alpha_{f}}\sigma\cos\theta,
\end{equation}
where $\alpha_f^{}=e^2/\hbar c\approx 1/137$ is the fine-structure constant. The
comparison of Equations~(\ref{eq:maxofen}) and (\ref{eq:maxclass}) shows that the former
depends on the strength and the curvature of the magnetic field,
whereas in Equation~(\ref{eq:maxclass}) the magnetic field only enters
through the ratio $\sigma\cos\theta=\mathcal{E}_{\parallel}/B$. In the case of a strong
electric field ($\sigma=1$, $\theta=0^{\circ}$) and the dominance of energy
losses by synchrotron radiation, the maximum energy of
radiation is restricted by $\epsilon_{syn}\approx 160$~MeV. If the energy losses
of electrons occur in the curvature or synchro-curvature
regimes, the position of the maximum moves to higher
energies. For example, for parameters used in our numerical
calculations, the radiation maximum can be as high as $30$ GeV 
(see Figure~\ref{fig:accel}). Thus, less intense energy losses of the
curvature radiation lead to more energetic radiation. Remarkably,
with an increase of curvature radius $r_0$ the radiation
becomes more energetic.

For comparison of the drift approximation and the exact
calculation, it is convenient, as above, to introduce the $q$-parameter.
However, in cases when only the magnetic field
presents, the $q$-parameter has been determined by Equation~(\ref{cur42})
as ratio of the curvature of the real trajectory and the curvature
of the magnetic field lines. Now, when we compare the real
trajectory and the \textit{electric} drift trajectory, we determine $q$-parameter
to be $q=R_{E}/R_{c}$, where $R_{E}$ and $R_{c}$ are the curvature
radii of the \textit{electric} drift and the real trajectory, respectively.
Then the characteristic frequencies of the radiation in both
cases are related as $\epsilon_{c}=q\epsilon_{E}$. One can see from the right panel
of Figure~\ref{fig:accel} that in the gap (green lines) the Lorentz factors are
close and the $q$-parameter is practically unity. The small
difference in the Lorentz factors produces a small difference in
the radiated spectra presented in the left panel of Figure~\ref{fig:accel} by
red and green lines. This difference could be explained by the
fact that the trajectories in the drift approximation and the exact
calculation are not identical, as can be seen from Figure~\ref{fig:traj}, and
the particle experiences slightly different magnetic and electric
fields.

When the particle escapes the outer gap it experiences the
decrease of the field-aligned electric field. We explore how this
drop in strength influences the radiation of the particle
depending on the intensity (gradient) of the drop. The relevant
Lorentz factors and $q$-parameter are shown in the right panel of
Figure~\ref{fig:accel}. In the case of the slow decrease of the field-aligned
electric field with $W=10^{-1}R_{lc}$ (pink lines), the $q$-parameter is
unity and the Lorentz factors for approximate and exact
calculations follow each other, gradually decreasing. In this
case the radiation spectra are close. For a more quickly
decreasing electric field ($W=10^{-2}R_{lc}$, blue lines) the $q$-parameter
oscillates around unity, and there is a slight
divergence between the Lorentz factors calculated in different
approaches (especially close to the gap border), but the
resulting spectra do not differ significantly. A different
situation occurs when the field-aligned electric field drops
very fast with screening width $W=10^{-3}R_{lc}$ (brown lines). The
$q$-parameter increases quickly on several orders of magnitude.
At the same time, the Lorentz factor in the exact calculation
drops, whereas in the drift approximation it continues to change
gradually. This produces different spectra calculated in two
approaches. The exact calculation gives more energetic
radiated photons compared to the drift approximation (solid
lines in the left panel of Figure~\ref{fig:accel}). Thus, in the case of fast
screening, the drift approximation does not work.

\section{Discussion}\label{sec:dis}
In curved magnetic fields the particle trajectories acquire a
complex structure when the curvature of the averaged drift
trajectory is superimposed on the regular curvature of its
gyration. The interplay of the two components produces a
variable curvature of the particle trajectory. This results in the
variability of radiation where the intensity and the spectrum are
changed with the period of the particle gyration. The variability
becomes pronounced during the transition from the synchrotron
to the curvature radiation regime when the two components of
the curvature of the trajectory become comparable. During this
transition the radiation cannot be described by synchrotron or
curvature modes, but rather by their combination -- synchro-curvature
radiation.

Because of the fast variability, only the averaged spectrum of
radiation can be observed. The spectrum averaged over the
period of gyration is described by Equation~(\ref{eq:59}) and depends
on two parameters. At small or large values of the $\eta$ parameter,
the spectrum coincides with the standard form of the
synchrotron spectrum. It is important to note that the radiation
could not be reduced to the formalism of the synchrotron
radiation by introducing a single effective curvature radius.
Indeed, the intensity of radiation averaged over the period of
gyration is given by Equation~(\ref{cur38}). It can be represented in the
standard form of Equation~(\ref{cur36}), if we introduce an effective
curvature radius $R_{\rm eff}=r_0/\sqrt{1+\eta^2}$. However, if we want to
have a standard exponential term e-x in an explicit form (as in
the case of the synchrotron or curvature radiation) in the
asymptotic presentation of the spectrum given by Equation~(\ref{asymp}), 
the effective curvature radius should be defined
differently, namely $R_{\rm eff}=r_0/(1+\eta)$.

The conditions for the realization of one or another regime
are determined by the relation between the drift velocity $\beta_D$ and
the velocity perpendicular to the drift trajectory $\beta_{\perp}$. The general
conditions of the applicability of synchro-curvature radiation
are also described in terms of $\beta_D$ and $\beta_{\perp}$ by Equation~(\ref{eq:applim}). In
the model case of a magnetic field of constant curvature and
without energy losses, $\beta_{\perp}$ is a free parameter. If $\beta_{\perp}=0$, the
particle can move along a drift trajectory without gyration. This
case is associated with the pure curvature radiation, which
occurs solely due to curvature of magnetic field line. In the
strong magnetic field, the drift trajectory is close to the field
lines, so the substitution of the curvature of the field lines
instead of the curvature of the drift trajectory is an acceptable
approximation. In particular, in the radiation formulas we have
neglected this difference assuming that the condition given by
Equation~(\ref{cur6}) is fulfilled.

The spectrum of radiation becomes quite sensitive to the
pitch angles if $\beta_{\perp}\geq\beta_D$, with a shift of the maximum to higher
energies roughly as $1+\eta$. Therefore any angular distribution
of particles wider than $\beta_D$ produces a superposition spectrum
with a high-energy cutoff that falls down slower than the
standard exponential cutoff of the synchrotron or curvature
radiation. Taking into account that in a strong magnetic field
$\beta_D$ is very small, even in the case of a very narrow angular
distribution, the radiation spectrum could differ significantly
from the unidirectional beam. This is demonstrated in Figures~\ref{f2}
and \ref{flat}, where the radiation spectra is averaged over the pitch-angles
of particles, assuming that Gaussian type and uniform
angular distributions of particles are presented. For the
Gaussian type angular distribution, the spectral flux density
can be presented in a simple analytical form by Equation~(\ref{flat}),
which provides good accuracy at $x \gtrsim 0.5$. As seen from
Figure~\ref{flat}, spectra becomes harder compared to the curvature
radiation ($\zeta=0$). It could also be noted that different angular
distributions with the equal width give similar spectra.

This interesting feature may be a key for the interpretation of
the recent observations of pulsars by {\it Fermi} LAT that indicate
that the energy spectra of some pulsars, in particular the Crab
pulsar \citep{Fermi}, in the cutoff region are
significantly harder than $e^{-x}$, as predicted by the curvature
radiation models. The energy spectrum of the pulsed emission
of the Crab reported in \citep{Fermi} can be readily
described by Equation~(\ref{dis3}) assuming $\zeta=1$ and a quite
reasonable value for the ratio $\gamma^3/r_0 \simeq 10^{13} \;{\rm cm}^{-1}$. It should be
noted that the form of the high-energy cutoff does not depend
on energy losses because it is produced by the most energetic
particles at the first moments of radiation before they lose their
energy. The angular distribution with enough broad angular
width could appear when the particles accelerated along
different field lines enter the region of the screened
electric field.

The results of the simplified analytical approach are
confirmed by detailed numerical calculations. More importantly,
the numerical calculations allow a rather general
treatment of the problem extending to the case when the
energy losses of electrons are taken into account. We studied
the radiation for the parameters typical for the polar cap and
outer gap models of the pulsar magnetosphere following the
evolution of the radiation of electrons with different initial pitch
angles in a dipole magnetic field. As expected, the least
energetic radiation is produced by electrons with an initial
angle along the drift trajectory. The particle radiates in the
synchro-curvature regime if the initial pitch angle is of the
order of $\beta_D$, in particular, if the initial direction is along the
magnetic field line. The spectra are substantially more energetic
when the initial pitch angle exceeds $\beta_D$. For typical parameters
of the polar cap model $\beta_D=2.2\cdot 10^{-9}(\gamma/10^{8})$, so even a
slight deflection from the field line results in the production of
synchrotron radiation in the quantum regime. Moreover, this
creates the interesting feature of a double peaked spectrum (see
Figure~\ref{fig:PCrad}) if the initial pitch angle is $1/\gamma_0$. 

In the context of the quantum synchrotron radiation, we
should note the following circumstance. The photons are
produced in a quantum regime when the pitch angle of the
charged particle is sufficiently large. The direction of the
produced photons is basically along the direction of the
charged particle. Thus, the photons are produced at large angles
to the magnetic field. The fact that the photons have
approximately the energy of the parent charged particles
implies that the parameter $\chi=B\gamma \sin\alpha/B_{cr} \geq 1$, thus the
photon should produce an electron-positron pair and initiate an
electromagnetic cascade. In this case, the cascade starts earlier
than in the case of gamma-rays produced along the magnetic
field lines.

Due to the variation of the strength and the curvature of the
dipole magnetic field, the change of radiation regimes is a quite
complex process as can be seen from the variation of the $q$-parameter
in Figures~\ref{fig:OGradComD}-\ref{fig:OGradComA}. The transition to the curvature
regime occurs with different rates for the polar cap and the
outer gap environment. Whereas in the polar cap the particle
makes a transition to the curvature regime very fast (see
Figure~\ref{fig:PC1g}), the particle in the outer gap can radiate in the
synchro-curvature regime most of the time while passing the
gap (see the second panel of Figure~\ref{fig:OGradComB}). Although in the polar
cap the synchrotron regime only lasts briefly, the particle loses
most of its energy during this regime. Namely, if the initial
pitch angle is $a/\gamma_0$ ($a\geq 1$), the initial Lorentz factor on the drift
trajectory is about $\gamma_0/a$. Since the typical Lorentz factor is
$\gamma\approx 10^{7}$, the huge energy losses could be caused by a very
small pitch angle.

The treatment of pitch angles is not self-consistent without
the consideration of particle acceleration. To study this
question we examined the model of the acceleration in the
electric and magnetic fields of the vacuum-retarded dipole
\citep{Deutsch1955}, as well as other simple models of the electric
field. The calculations show that the particle basically moves
along the drift trajectory jointly defined by the electric and
magnetic fields. Therefore the approach of drift approximation
applied in the work of \cite{Kalapotharakos2012} gives
results that are similar to those obtained in the exact
consideration. The difference could arise when the particle
escapes the acceleration gap experiencing the decrease of the
field-aligned electric field. It has been shown that in the case of
the fast screening of the electric field the particle can start to
radiate more intensely. As a result, the particle radiates more
energetic photons than is predicted by the drift approximation.
In general, the drift approximation and the curvature radiation
can only be applied to the equilibrium situations when the
model parameters change slowly.

\appendix
\section{Motion of a charged particle in electric and magnetic fields}\label{sec:accel}
The motion of an ultrarelativistic charged particle in the
electric ($\mathbfcal{E}$) and magnetic ($\b B$) fields is described by the system
of equations \citep{Landau2}:
\begin{align}\label{eq:system}
&\frac{d \b r}{dt}=c\b \beta, \\\nonumber
&\frac{d \b \beta}{dt}=\frac{e}{mc\gamma}(\mathbfcal{E}-(\b \beta\mathbfcal{E})\b \beta+ \b\beta\times \b B), \\ \nonumber
&\frac{d\gamma}{dt}=\frac{e}{mc}(\b\beta\mathbfcal{E})-\frac{2e^2}{3mc}\frac{\gamma^4}{R_c^2},
\end{align}
where the curvature radius is expressed as
\begin{equation}\label{eq:curv}
\frac{1}{R_c}=\left| \frac{d \b \beta}{c dt} \right|=\frac{e}{mc^2\gamma}
\sqrt{(\mathbfcal{E}+\b\beta\times\b B)^2-(\b \beta \mathbfcal{E})^2}. 
\end{equation}
The motion of an ultrarelativistic charged particle in the  {\it drift approximation} is determined by the
following system of equations:
\begin{align}\label{eq:systemApp}
&\frac{d \b r}{dt}=c\b \beta, \\\nonumber
&\frac{d\gamma}{dt}=\frac{e}{mc}(\b\beta\mathbfcal{E})-\frac{2e^2}{3mc}\frac{\gamma^4}{R_c^2},
\end{align}
where
\begin{equation}
\b\beta=\frac{\mathbfcal{E} \times \b B}{B^2}+f\frac{\b B}{B},
\end{equation}
and $f$ is defined from the condition that $\left|\b \beta \right|=1$
\begin{equation}
f=\sqrt{1-\left(\frac{\mathbfcal{E} \times \b B}{B^2}\right)^2}.
\end{equation}
Then the curvature of the trajectory is
\begin{equation}
\frac{1}{R_c}=\left|\b K\right|=\left|(\b \beta\nabla)\b \beta\right|.
\end{equation}
Note that the differential equations that describe the evolution of Lorentz factor have the same  form
in both approaches. The difference is in the determination of the curvature radius. 
\section{Energy losses in the quantum regime}\label{sec:appB}
In a strong magnetic field,  the  ultrarelativistic electrons  can radiate in the quantum regime, 
provided that
\begin{equation}
\chi=\frac{B}{B_{cr}}\gamma\sin\alpha \gtrsim 1,
\end{equation}
where $B_{cr}=\frac{2m^2c^3}{3e\hbar}\approx 2.94\cdot 10^{13}$ G.
In the presence of electric field, $B\sin\alpha=|(\b\beta\times\b B)|$ should be substituted by
$\sqrt{(\mathbfcal{E}+\b\beta\times\b B)^2-(\b \beta \mathbfcal{E})^2}$. The energy loss rate 
can be written in the form \citep{Bayer,Landau4}
\begin{equation}\label{eq:QEnLoss}
\left| \frac{dE}{dt} \right|=\frac{e^2 m^2 c^3}{\sqrt{3}\pi\hbar^2} \overline{H}(\chi) \, ,  
\end{equation}
where 
\begin{equation}
\overline{H}(\chi)=\int_{0}^1 H(\tau,\chi) d\tau \ ,
\end{equation}
and  
\begin{eqnarray}
H(\tau,\chi)=\chi\left[(1-\tau)F(x)+x\tau^2K_{2/3}(x) \right],
\quad x=\frac{\tau}{\chi(1-\tau)},
\end{eqnarray}
where $K_{2/3}(x)$ is the modified Bessel function of the order $2/3$, $F(x)$ is
the emissivity function of the synchrotron radiation (see Equation~\ref{cur41}), $\tau=\epsilon/E$, where
$\epsilon$ is the energy of the radiated photon, $E$ is the energy of the radiating particle.

For  calculations it is  convenient to express  $\overline{H}(\chi)$ in Equation~(\ref{eq:QEnLoss}) in 
a simple  approximate analytical  form. Using asymptotics of this  function
\begin{eqnarray}
\begin{aligned}
&\overline{H}(\chi)\approx {\frac {8\pi\sqrt {3}}{27}}{\chi}^{2},\quad  \chi\ll 1,  \\
&\overline{H}(\chi)\approx {\frac {32\pi\sqrt {3}}{243}}{2}^{2/3}\Gamma\!
\left(\frac{2}{3} \right) \chi^{2/3},\quad
\chi\gg 1.
\end{aligned}
\end{eqnarray}
we suggest the following approximation
\begin{eqnarray}\label{eq:QLossApp}
\overline{H}(\chi)\approx\frac{8\pi\sqrt{3}}{27}\frac{\chi^2}{\left({\displaystyle 1+\frac{3}{4}\frac{(2\chi)^{2/3}}
{\sqrt{\Gamma\!\left(\frac{2}{3} \right) }}} \right)^2 }
\times \left(1+\frac{0.52\sqrt{\chi}(1+3\sqrt{\chi}-3.2\chi)}{1+0.3\sqrt{\chi}+17\chi+11\chi^2}\right)
\end{eqnarray}
The first part of Equation~(\ref{eq:QLossApp}) (before the sign $\times$)  gives right asymptotics at $\chi\ll 1$ and $\chi \gg 1$ and
provide an  accuracy better than $10\%$ for other values of $\chi$, whereas the last term in the brackets
improves the accuracy down to $0.1\%$ for any  $\chi$.

The spectrum of radiation in the quantum regime is described by the function that can be written as
\begin{eqnarray}\label{eq:fq}
F_q(x,\tau)&=(1-\tau)F(x)+\tau^2 x K_{2/3}(x),
\quad x&=\frac{\tau}{1-\tau}\frac{E}{\epsilon_c},
\end{eqnarray}
where $\epsilon_c=\frac{3e\hbar B\sin\alpha}{2 mc}\gamma^2$ is the characteristic  energy of the emitted photon.
To use this function in Equation~(\ref{cur43}), the variable $\epsilon_c$ should be replaced by $\hbar\omega_{*}$.
An  analytical approximation of this function can be obtained using the approximation for emissivity function of
the synchrotron radiation \citep{Aharonian2010}
\begin{eqnarray}\label{eq:synchapp}
F(x)\approx 2.15 x^{1/3}(1+3.06 x)^{1/6}
\frac{1+0.884x^{2/3}+0.471 x^{4/3}}{1+1.64 x^{2/3}+0.974 x^{4/3}}e^{-x},
\end{eqnarray}
and
\begin{eqnarray}
x K_{2/3}(x)\approx 1.075 x^{1/3}(1+3.72 x)^{1/6}
\frac{1+1.58x^{2/3}+3.97 x^{4/3}}{1+1.53 x^{2/3}+4.25 x^{4/3}}e^{-x}.
\end{eqnarray}
Both approximations provide an accuracy better than $0.2\%$  for any 
value of the argument $x$.
\section{Maximum of the curvature radiation spectrum}\label{sec:maxen}

The maximum energy of the electron determined from the competition between the acceleration and the energy
losses, is found from the condition $t_{acc}=t_{loss}$, where
\begin{equation}
t_{loss}=\frac{E}{|dE/dt|_{loss}}=\frac{3mcR_c^2}{2e^2\gamma^3},
\end{equation}
and
\begin{equation}
t_{acc}=\frac{E}{|dE/dt|_{acc}}=\frac{E}{ce\sigma \cos\theta B}=\frac{mc\gamma}{e\sigma\cos\theta B}.
\end{equation}
Note that the acceleration rate is defined as $|dE/dt|_{acc}=ce\mathcal{E}\cos\theta$, where $\mathcal{E}$ is electric
field, $\sigma=\mathcal{E}/B$ is the ratio of the  electric and magnetic fields, and $\theta$ is the angle between them. From $t_{loss}=t_{acc}$ we obtain Lorentz factor of the electron:
\begin{equation}
\gamma_{eq}=\left(\frac{3\sigma \cos\theta B R_c^2}{2e}\right)^{\frac{1}{4}}.
\end{equation}
Substituting this equation into the expression for the characteristic energy of radiation given
by Equation~(\ref{cur40}) we find
\begin{equation}\label{eq:maxen}
\epsilon_{*}=\left(\frac{3}{2}\right)^{\frac{7}{4}}\left(\frac{\sigma\cos\theta B}{e}\right)^{\frac{3}{4}}c\hbar\sqrt{R_c
}.
\end{equation}
Finally, taking into account that the curvature of the drift trajectory is $R_c=r_0(1+\beta_{E}^2)$,
where $\beta_{E}=\sin \alpha$ is the drift velocity due to the electrical drift, and $\alpha$ is the pitch
angle given by Equation~(\ref{eq:electdr}), we obtain
\begin{equation}
\epsilon_{*}=\left(\frac{3}{2}\right)^{\frac{7}{4}}\left(\frac{\sigma\cos\theta B}{e}\right)^{\frac{3}{4}}c\hbar\sqrt{r_0
\left(1+\sigma^2\sin^2\theta\right)},
\end{equation}
where $r_0$ is the curvature of the magnetic field line.

\end{document}